\documentclass[11pt,a4paper]{article}

\usepackage[T1]{fontenc}
\usepackage[latin2]{inputenc}
\usepackage{graphicx}
\usepackage[normalem]{ulem}

\usepackage{fancyhdr}
\usepackage{calc}
\usepackage{epsfig}
\usepackage{amssymb}
\usepackage{amsmath}
\usepackage{indentfirst}
\usepackage{chngcntr}
\usepackage{jheppub}

\def\s0{\sigma_0}

\title{Twist decomposition of Drell-Yan structure functions: phenomenological implications}
\author[a,b]{Dawid Brzemi\'{n}ski,}
\author[a]{Leszek Motyka,}
\author[a]{Mariusz Sadzikowski}
\author[a]{and Tomasz Stebel}
\affiliation[a]{Institute of Physics, Jagiellonian University,\\
S.\ \L{}ojasiewicza 11, 30-348 Krak\'{o}w, Poland}
\affiliation[b]{Cavendish Laboratory, University of Cambridge,\\ 19 J.J. Thomson Avenue, Cambridge CB3 0HE, U.K.}
\emailAdd{dawid.brzeminski@gmail.com}
\emailAdd{leszek.motyka@uj.edu.pl}
\emailAdd{mariusz.sadzikowski@uj.edu.pl}
\emailAdd{tomasz.stebel@uj.edu.pl}
\abstract{
The forward Drell--Yan process in $pp$ scattering at the LHC at $\sqrt{S}=14$~TeV is considered. We analyze the Drell--Yan structure functions assuming the dominance of a Compton-like emission of a virtual photon from a fast quark scattering off the small~$x$ gluons. The color dipole framework is applied to perform quantitatively the twist decomposition of all the Drell--Yan structure functions. Two models of the color dipole scattering are applied: the Golec-Biernat--W\"{u}sthoff model and the dipole cross section obtained from the Balitsky--Fadin--Kuraev--Lipatov evolution equation. The two models have essentially different higher twist content and the gluon transverse momentum distribution and lead to different significant effects beyond the collinear leading twist description. It is found that the gluon transverse momentum effects are significant in the Drell--Yan structure functions for all Drell--Yan pair masses $M$, and the higher twist effects become important for $M \lesssim 10$~GeV. It is found that the structure function $W_{TT}$ related to the $A_2$ angular coefficient and the Lam--Tung observable $A_0 -A_2$ are particularly sensitive to the gluon $k_T$ effects and to the higher twist effects. A procedure is suggested how to disentangle the higher twist effects from the gluon transverse momentum effects. 
}
\begin{document}

\maketitle

\section{Introduction}

The Drell--Yan process is a classical probe of the proton structure and of the strong interactions in hadron collisions \cite{DrellYan}. The experiments operating at the Large Hadron Collider have already detected large statistic of the Drell--Yan dileptons and measured the differential Drell--Yan cross sections as functions of several kinematical variables \cite{DYgamma_atlas,CMS:2014jea,Aad:2014xaa,Khachatryan:2015oaa,CMSLT,Aaij:2015gna,Aaij:2015zlq,ATLASZ0}. In particular the angular distributions of the dileptons were measured at the $Z^0$ peak that allow the determination of the Drell--Yan structure functions at the mass close to the $Z^0$ boson mass \cite{CMSLT,ATLASZ0}. The Drell--Yan measurements when extended to the low mass region, $M < 10$~GeV, may be used to provide unique information about parton densities in the proton at a very low~$x$, at or below $x=10^{-5}$ \cite{lhcb0}. In this kinematic region of the low $M$ and very small~$x$ the higher twist corrections may affect significantly the Drell--Yan cross section and hence the parton density function determination from the data, see e.g.\ Ref.\ \cite{GBLS,MoSadSte}. Therefore, in order to achieve the highest precision of the parton density function determination the higher twist corrections should be taken into account in the analysis. Unfortunately not much is known about the higher twist contributions to the proton structure. The subleading twist~4 corrections are represented by a set of independent operators whose matrix elements have not been measured yet, see e.g.\ Refs.\ \cite{BFKL2,BraunTwist}. Therefore in the estimates of the higher twist corrections to proton scattering cross sections it is still necessary to relay on models. An approach to model of the higher twist effects in high energy scattering at small~$x$ was proposed on the basis of the Golec-Biernat--W\"{u}sthoff (GBW) saturation model \cite{GBW}, that provides an efficient unified picture of the high energy scattering down to very low scales where multiple scattering and higher twist effects are expected to contribute. The framework for extraction of the twist components from the GBW model was formulated for the DIS at HERA in \cite{BGBP,BGBM}, then further developed and applied to the diffractive DIS at HERA \cite{MoSadSlo} and the forward Drell--Yan cross sections \cite{GBLS,MoSadSte}. Within this framework an evidence of the higher twist corrections to DDIS structure functions was found in \cite{MoSadSlo} and recently, a related approach revealed an evidence of the higher twist corrections to the proton structure functions at small~$x$ \cite{HERAtwist}.

A framework that is capable to provide a QCD guideline for extending theoretical analysis beyond the twist~2 collinear approach, is the $k_T$ factorization formalism \cite{GLR,bfkl,bfklrev,CCH,CH}. The treatment of the forward Drell--Yan process within the $k_T$ factorization was initially proposed in \cite{BroHeb} in the color dipole representation \cite{NZ}, and then further developed and applied to data analysis in numerous papers, see e.g.\ Refs.\ \cite{Kopeliovich,Kopeliovich:2001hf,GJ,Ducati,VG}. Later on also the momentum representation of the forward and general Drell--Yan process were elaborated in detail \cite{SchSz,MoSadSteLT}. The dipole formulation of the forward Drell--Yan scattering was used to obtain the twist decomposition of the Drell--Yan cross section integrated over the dilepton angular distribution \cite{GBLS}. In the latter analysis the GBW of the QCD dipole cross section was assumed. In a recent paper \cite{MoSadSte} we prepared the theoretical framework to extend this type of twist analysis to all the Drell--Yan structure functions. We also discuss there in more detail earlier estimates of the higher twist content of the Drell--Yan cross section performed in Refs.\ \cite{Berger,QS,Brandenburg,Eskola,FMSS,FSSM,BMS}.

In the present paper we apply the results of our earlier paper \cite{MoSadSte} to perform quantitative estimates of the higher twist contributions to the Drell--Yan structure functions based on the GBW saturation model \cite{GBW}. Moreover we derive the analytic formulae for the twist decomposition of the forward DY structure functions assuming the Balitsky--Fadin--Kurayev--Lipatov (BFKL) pomeron exchange \cite{bfkl,bfklrev}. The two approaches assume an essentially different dynamics of multiple hard scattering and have an essentially different twist content. In the GBW model a simple eikonal picture of multiple scattering is applied corresponding to the resummation of independent single exchanges. This leads to the higher twist amplitudes strongly enhanced by inverse powers of $x$ at small~$x$. The BFKL pomeron exchange amplitude emerges as a QCD result obtained from the resummation of the leading logarithms of~$x$. The BFKL pomeron exchange implicitly carries higher twist contributions which, however, are power suppressed by the positive powers of $x$ at small~$x$~\cite{MoSad} in striking contrast to the eikonal picture. Also, in the BFKL amplitude the multiple gluon ladder exchanges leading to higher twist contributions are correlated, whereas they are not correlated in the eikonal picture. 

The two considered pictures of multiple gluon exchange differ also in the $k_T$ shapes of the gluon transverse momentum distribution (TMD). The GBW model leads to a narrow, quasi-collinear gluon $k_T$ distribution with the width scale given by the saturation scale, that is ${\cal O}(1~\mathrm{GeV})$, and the BFKL evolution generates a wide, power-like transverse momentum distribution with the asymptotic (at a very small $x$) positive anomalous dimension of $1/2$, leading to $\sim 1/k_T$ behavior of the gluon TMD ${\cal F}(x,k_T^2)$. Since both the initial parton $k_T$ and the higher twist effects influence the Drell--Yan structure functions, it is desirable to analyze and disentangle them. The two considered models are particularly useful for this purpose as the GBW model introduces sizable higher twist effects and very small gluon $k_T$, whereas the BFKL exchange generates gluons with large $k_T$ but it leads to very small higher twist corrections at small~$x$.

The color dipole description of the Drell--Yan process incorporates small~$x$ resummation effects and a multiple scattering resummation (within a model). In particular the Drell--Yan description with BFKL amplitudes may be related to an analysis of small~$x$ effects in the DY scattering performed in Ref.\ \cite{Marzani:2008uh}. The Drell--Yan process, however, receives large perturbative QCD corrections also in the limit of $q_T \ll M$ and from soft gluon radiation near the partonic threshold energy. They are not included in the standard form of the dipole models. In particular the corrections coming from the small $q_T$ region
of the cross section introduce at all orders $n$ of the perturbative expansion terms enhanced by double logarithms $\sim \alpha_s ^n \log^{2n-1} (q_T ^2 / M^2)$, and also subleading logarithmic corrections \cite{CSS}. Furthermore, it was shown by Collins, Sterman and Soper (CSS) \cite{CSS} that in the relevant small $q_T$ region these corrections may be resummed or parameterized by a universal non-perturbative transverse momentum dependent parton distribution at very small $q_T$. It was proven that the contribution of the resummed corrections of this type cancel after $q_T$ integrations \cite{Collins:2016hqq}, but it is essential for the correct description of the Drell--Yan $q_T$-dependent cross section at small $q_T$. Recently the problem of joint resummation of small~$x$ effects and the transverse momentum logarithms was addressed \cite{Forte:2015gve,Marzani:2015oyb} providing important results for analyses of the $q_T$ dependent DY distribution at small~$x$. The scheme for the joint resummations of transverse momentum logarithms and threshold correction is also available (see e.g.\ \cite{Kulesza:2002rh,Lustermans:2016nvk,Marzani:2016smx}) but it is expected to have less impact on the small~$x$ cross section. To summarize, the dipole model and BFKL predictions for the $q_T$~dependent DY cross sections need an improvement by the CSS resummation but the effects of this resummation cancel in the $q_T$ integrated cross section.

The paper is organized as follows. In section \ref{strFunc} one can find basic definitions of the Drell--Yan structure functions $W_j$ and of two models, the BFKL exchange, and the GBW saturation model. In the next section the procedure of the twist decomposition is discussed. The numerical predictions can be found in section \ref{results_unint}. These predictions are presented in terms of the dimensionless structure functions $A_i$ commonly used for data presentations. Section \ref{twist_exp_int} contains definitions of the structure functions $\tilde{W}_i$ integrated over the lepton pair transverse momenta. It also contains a discussion of their twist decomposition which requires some attention due to apparent singularities. In section \ref{results_int} numerical results are presented in terms of the $\tilde{A}_i$ and invariant $\lambda_i$ structure functions. The conclusions are given in section \ref{concl}.

\section{Structure functions in Drell--Yan processes}
\label{strFunc}

The DY helicity structure functions $W_j$ are defined through the formula for the differential cross section \cite{LamTung1,LamTung2}
\begin{eqnarray}
\frac{d\sigma}{d x_F dM^2 d \Omega d^2 q_T} & = & \frac{\alpha^2_{em}}{2(2\pi)^4 M^4} \left[ (1-\cos ^2 \theta) W_L + (1+\cos ^2 \theta) W_T
+ (\sin^2\theta \cos 2\phi)W_{TT} \right. \nonumber\\
& + & \left. (\sin2\theta \cos \phi) W_{LT}\right],
\label{sigma_diff_W}
\end{eqnarray}
where $M$ is the lepton pair invariant mass, $q_T$ --- the transverse momentum of the virtual photon and $x_F$ --- its Feynman parameter. $(\theta,\phi)$ are the polar and azimuthal angles of the lepton momentum vector in the dilepton c.m.s.\ frame. The frame orientation is not unique, and the most common frame choices are the Collins--Soper frame \cite{CSframe} and the Gottfried--Jackson frame \cite{GJframe}.
In this paper we apply the description of the forward Drell--Yan process in the color dipole formulation. This formulation \cite{BroHeb} assumes the dominance of the Compton-like partonic channel in which the fast collinear quark $q$, coming from one of the protons scatters off the color field of the other proton by single or multiple virtual gluon $g^*$ exchanges. At the lowest order the perturbative partonic channel for the forward Drell--Yan process with an intermediate $\gamma^*$ is $qg^* \to q'\gamma^* \to q' l^- l^+$, where $l^-$ and $l^+$ denote the produced leptons. The detailed description of the kinematics, and the relevant diagrams may be found in Ref.\ \cite{MoSadSte}. Hence, in the $k_T$ factorization approach within the color dipole picture defined in \cite{BroHeb,Kopeliovich} one can show that in the Gottfried--Jackson frame the DY structure functions may be expressed as,
\begin{eqnarray}
W_{j}=\int_{x_F}^1 dz \ \wp(x_F/z)
\int_{\cal C} \frac{ds}{2\pi i} \ \tilde{\sigma} (s) \left ( \frac{z^2 Q_0^2}{M^2(1-z)} \right) ^{-s} \hat{\Phi}_{j} (q_T,-s,z),
\label{Wphihat}
\end{eqnarray}
where $z$ is the longitudinal momentum fraction of the initial state quark taken by the virtual photon, $\wp(x_F/z)$ is a collinear parton distribution function and $\tilde{\sigma} (s)$ is a color dipole -- proton cross section in the Mellin representation. The parameter $Q_0$ is the Mellin transform scale. The leptonic impact factors $\hat{\Phi}_{i}$ calculated in \cite{MoSadSte} can be found in Appendix \ref{impact_facts_unint}. The integration contour in the complex plane is taken as $\mathcal{C} = (-1/2-i \infty, -1/2 + i \infty)$. 

We consider two models for the description of the color dipole cross section:
%\begin{itemize}
%\item 
%
\paragraph{1.\ The GBW model} \cite{GBW} in the form:
\begin{equation}
\label{GBW_param} 
\sigma (\rho) = \sigma_0 (1-e^{-\rho^2}), \;\;\mbox{where}\;\; \rho = rQ_0(x)/2,
\end{equation}
where $Q_0$ is the saturation scale, $Q^2 _0(x) = (\hat x_0 / x)^{\lambda}$~GeV$^2$. The Mellin transform of the GBW dipole cross section w.r.t.\ the $\rho^2$ is equal to $\tilde{\sigma}_{GBW} (s) = -\sigma_0\Gamma(s)$. The parameter values of the original GBW model \cite{GBW} are applied: $\lambda =0.288 $, $\sigma_0=23.03$~mb, and the value of  $\hat x_0 = 6.08 \cdot 10^{-4} $ is chosen to be two times larger than the original GBW value $x_0 = 3.04 \cdot 10^{-4}$ obtained from the description of the DIS data \cite{GBW}. This modification was introduced because for the DIS data description the $x$ variable in the dipole cross section was set to the threshold value of the gluon $x_g$, that is to $x=Q^2/W^2$ (where $W$ in the proton--$\gamma^*$ collision energy), while in our DY description with $q_T$-dependence the gluon $x_g$ is derived from the exact kinematics instead of using its threshold value, see Eq.\ \ref{xg_def}. To be more specific, in the $k_T$-factorization picture of the DIS at small~$x$ with the exact kinematics, the value of gluon~$x_g$ in the $\gamma^* g \to q\bar q$ process depends on the virtual photon $Q^2$ and on the mass $M_X$ of the produced $q\bar q$ state: $x_g \simeq (Q^2 + M_X^2)/W^2$, (see e.g.\ \cite{Kwiecinski:1997ee} for the detailed discussion). Since the typical mass of the produced partonic $q\bar q$ state from the $\gamma^*$ fragmentation is $M_X \sim Q$, the approximate value of the gluon $x_g$ in the DIS is significantly larger than its threshold value $Q^2/W^2$ and may be approximately estimated as $x_g \simeq 2\,Q^2/W^2 = 2x$. So, if one treats the GBW dipole cross section as a function of true gluon $x_g$ instead of its threshold limit $x$, as we do for the $q_T$-dependent DY cross section, it is necessary to rescale the model parameter $x_0$ to approximately $2x_0$. Then at given values of the observed parameters $Q^2$ and $W^2$, the dipole cross section expressed through $x_g \simeq 2x$ and the rescaled parameter $2x_0$ is the same as the dipole cross section expressed through $x$ and $x_0$. Our default choice for the figures in the paper is the parameter $\hat x_0$, but as the described treatment of the gluon kinematics is only approximate, we shall also explicitly display the sensitivity of selected observables to the choice of between the original GBW value $x_0$ and the rescaled value $\hat x_0$ of the dipole cross section. 

\paragraph{2.\ The BFKL dipole cross section} based on the solution of the leading-order (LO) BFKL equation~\cite{bfklrev} with
the GBW input at a chosen value of $x$: $x_{\textrm{in}}= 0.1$. The solution in the Mellin space reads
\begin{equation}
\label{sigma_BFKL}
\tilde{\sigma}_{BFKL}(s,Y) = -\sigma_0'\Gamma(s) e^{\bar{\alpha}_s\chi(s)Y}.
\end{equation}
where $\chi(s)$ is the LO BFKL characteristic function
\begin{equation}
\chi(s) = 2\psi(1)-\psi(-s)-\psi(1+s) ,
\end{equation}
expressed through the digamma function $\psi$.
The cross section parameter $\sigma_0' = 2\pi R_p^2$ where $R_p$ is an effective radius of the proton which emerges after integration of an imaginary part of the forward dipole-nucleon scattering amplitude over the impact parameter $\mathbf{b}$. The rapidity evolution length $Y$ is given by
\begin{equation}
\label{Y_def}
Y=\log \left( \frac{x_{\textrm{in}}}{x_{g}} \right),
\end{equation}
where the value of gluon $x_{g}$ follows from the kinematics of the forward Drell--Yan process in the $qg^* \to q\gamma^*$ channel:
\begin{equation}
\label{xg_def}
x_{g} = \frac{(1-z) M^{2}+q_{T}^{2}}{S\, x_{F} (1-z)},
\end{equation}
with $S$ denoting the invariant mass squared of the pair of colliding proton beams.
In equation (\ref{sigma_BFKL}) we adopted the eikonal form of the initial condition for the BFKL evolution, coming the Golec-Biernat--W\"usthoff (GBW) model \cite{GBW}:
\begin{equation}
\label{in_cond}
\sigma_{BFKL}(r,Y=0) = \sigma_0' (1-e^{-r^2 \bar Q_0^2 /4}) .
\end{equation}
The model parameters were set by the fit to the DIS data \cite{MoSad} and read $\bar{\alpha}_s = 0.087$, 
$\bar Q_0 = 0.51$ GeV, $\sigma_0' = 17.04$~mb.

\begin{figure}
\centering
\includegraphics[width=.69\textwidth]{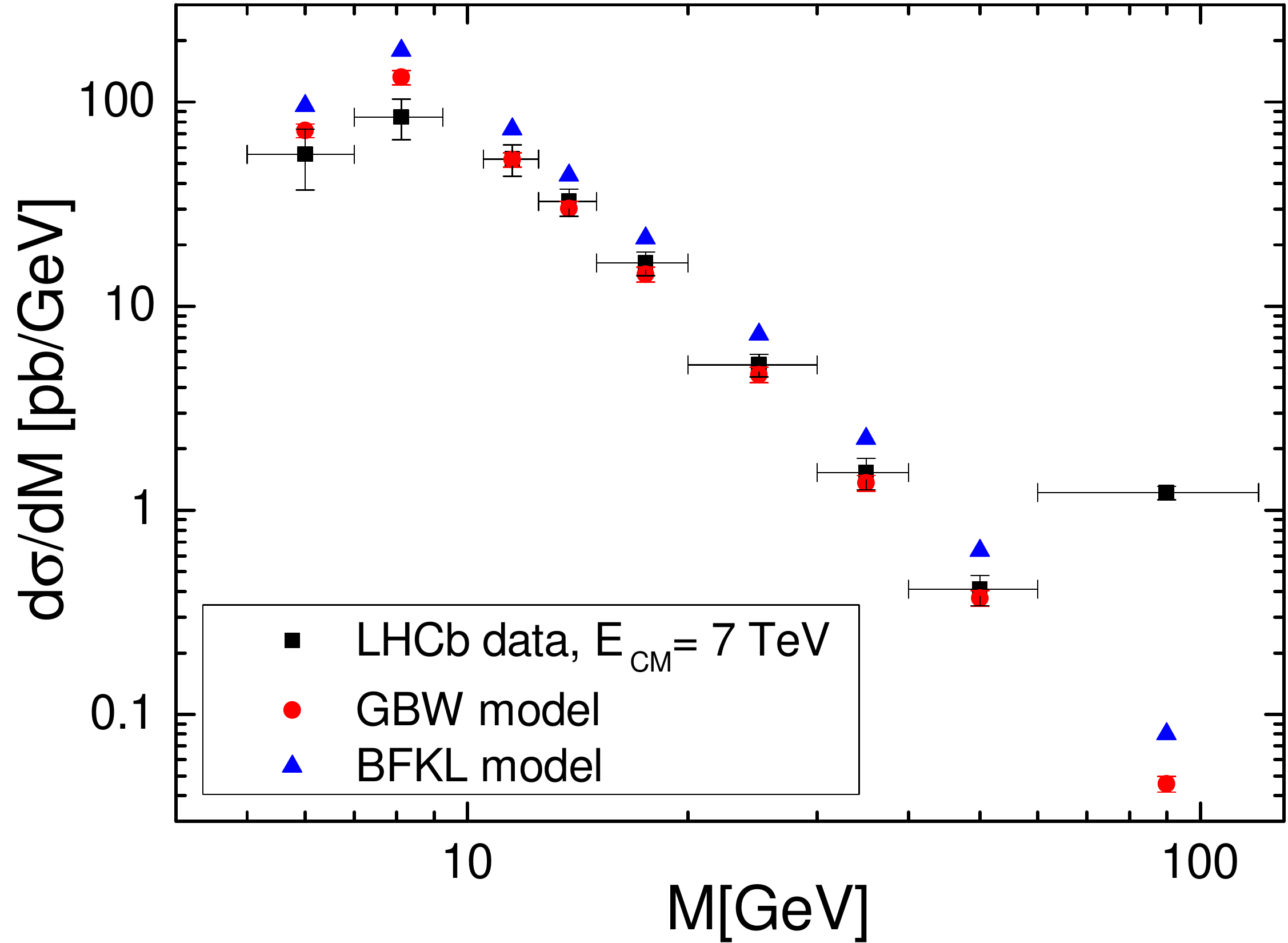}
\caption{Predictions of the GBW and BFKL models compared to data for the total DY cross section from LHCb experiment \cite{LHCb_data}. Horizontal error bars represent the bin sizes. The vertical error bars for the GBW model (within the point size) reflect the sensitivity of the predictions to the choice of the dipole model parameter $x_0$, see the discussion below Eq.\ (\ref{GBW_param}). 
}
\label{totalxsection}
\end{figure}

In our analysis we shall perform the twist decomposition of the DY structure functions obtained with the two model cross sections. Prior to that however, we test the reliability of the description by comparing the predictions for the forward Drell--Yan cross sections integrated over the lepton angles to the LHCb data. The results are shown in Fig.\ \ref{totalxsection} as functions of the DY pair mass~$M$ for both dipole cross section models. In the cross section calculations the kinematical cuts were taken from \cite{LHCb_data} with an approximate treatment of the lepton transverse momentum (to be precise --- when we impose the LHCb cuts on the lepton transverse momenta, $q_T$ is neglected in the transverse momentum balance of the leptons. This has a negligible effect on the results for $M>10$~GeV). Note that in the region of the $Z^0$ mass, at $M \simeq 90$~GeV the Drell--Yan cross section is dominated by the $Z^0$ boson production that is not included in our analysis, hence the data point at $M=90$~GeV is not well described. Except of this region, and for $M>10$~GeV where the approximate treatment of the cuts may be neglected, the GBW agrees well with the data. The theoretical uncertainty of the GBW predictions due to the choice of the $x_0$ parameter of the dipole cross section is indicated as the vertical error bars, which are however within the GBW point size for all the points. The upper values correspond to $\hat x_0$ and the lower ones to the standard GBW $x_0$. As seen from the figure, the sensitivity to the choice of $x_0$ is found to be small. The $M$~shape obtained with the BFKL model is consistent with the data but the overall normalization is slightly overestimated. The overall normalization, however, cancels in the analysis of the relative twist content so the small discrepancy of this parameter does not prohibit using this model in the twist analysis. Note that in the $q_T$ integrated cross sections the effects of the CSS resummation cancel \cite{Collins:2016hqq}.

\section{Twist expansion for the helicity structure functions}
\label{twExp}

The forward Drell--Yan helicity structure functions (\ref{Wphihat}) can be written as
\begin{equation}
\label{structure func redefined}
W_{j} = \int\limits_{x_F}^{1}dz \, \wp(x_{F}/z) \, \sigma_{j}(q_{T},z,Y),
\end{equation}
where
\begin{equation}
\label{structure dis analogue}
\sigma_{j}(q_{T},z,Y) = \int\limits_{\mathcal{C}}\frac{ds}{2 \pi i} 
\left(\frac{z^{2} \bar Q_{0}^{2}}{M^2(1-z)}\right)^{-s}
\tilde{\sigma}(s,Y) \hat{\Phi}_{j}(q_{T},-s,z) .
\end{equation}
The twist analysis of the forward DY cross section assuming the GBW dipole cross section was performed analytically in the preceding paper \cite{MoSadSte}. 

Below we perform an analogous twist decomposition using the BFKL model of the color dipole cross section using the method proposed in \cite{MoSad}. Hence, in order to perform the twist decomposition, we close the contour $\mathcal{C}$ with a left semicircle without changing the value of the integral. The integral over the closed contour is proportional to the sum of residues at the enclosed singularities. Hence, we express this integral as a sum of integrals around the singularities, which in this case are at the negative integers. The singularity at $s=-n$ is identified with the twist-$2n$ contribution to the amplitude. Therefore the cross section may be decomposed in the following way:
\begin{equation}
\sigma_{j}(q_{T},z,Y) = \sum_{n = 1}^{\infty} \sigma_{j}^{(2n)}(q_{T},z,Y),
\end{equation}
where $2n$ corresponds to twist-$2n$ term and singularity in $s=-n$.
If the BFKL dipole cross section is assumed then essential singularities appear at $s=-n$. A procedure to evaluate the corresponding residues was described in \cite{MoSad}. The essential singularities are enclosed by circles with the radius $\epsilon \to 0$, and the angle $\theta$ parameterizes the position on the circle corresponding to the singularity at $s=-n$. 
\begin{equation}
\label{sigma2n}
\sigma_{j}^{(2n)}(q_{T},z,Y)= - R_{p}^{2} e^{-nt}\int\limits_{0}^{2 \pi} 
d\theta  h^{(2n)}_{j}(\epsilon e^{i\theta},q_T,z,Y) \exp\left(\epsilon e^{i\theta}\, t  + \frac{\bar{\alpha}_{s}Y}{\epsilon}e^{-i\theta}\right),
\end{equation}
where
\begin{equation}
h^{(2n)}_{j}(\epsilon e^{i\theta},q_T,z,Y) \, = \, \epsilon e^{i\theta}  \left(\frac{z^2}{1-z}\right)^{n-\epsilon \exp{i \theta}} 
 \hat{\Phi}_{j}(q_{T},n-\epsilon e^{i\theta},z) \, 
\Gamma(-n+\epsilon e^{i\theta}) \, e^{\bar{\alpha}_{s} Y \chi^{(n)}_{reg}},
\end{equation}
and
\begin{equation}
\chi_{reg}^{(n)} = \chi\left(-n+\epsilon e^{i\theta}\right) - \frac{e^{-i\theta}}{\epsilon},
\end{equation}
is a regular function of $\epsilon$ in the limit of $\epsilon\rightarrow 0$ and $t=\log(M^2/\bar Q_0^2)$. Note that the terms which generate the essential singularities of the BFKL cross section coming from the exponentiated poles $\sim 1/\epsilon$ of the BFKL characteristic function were explicitly isolated. The coefficients $h_{j}^{(2n)}$ can be expanded into an infinite series in $\epsilon$,
\begin{equation}
\label{hn2}
h^{(2n)}_{j}(\epsilon e^{i\theta},q_T,z,Y)
= \sum_{m = 0}^{\infty} a_{m}^{(2n)j} \left(\epsilon e^{i\theta}\right)^{m},
\end{equation}
where the arguments $q_T$, $z$, and $Y$ of the series coefficients $a_{m}^{(2n)j}$ are suppressed 
After substitution of (\ref{hn2}) into (\ref{sigma2n}) and integration over the angle $\theta$ one gets,
\begin{equation}
\sigma_{j}^{(2n)}(q_{T},z,Y)= -2 \pi R_{p}^{2} \left(\frac{ \bar Q_{0}^{2}}{M^{2}}\right)^{n}\sum_{m = 0}^{\infty}a_{m}^{(2n)j} \left(\frac{\bar{\alpha}_{s}Y}{t}\right)^{\frac{m}{2}} I_{m} \left(2\sqrt{\bar{\alpha}_{s}Y t}\right) ,
\end{equation}
what combined with \eqref{structure func redefined} gives
\begin{equation}\label{structure func redefined2}
W_{j}^{(2n)} =-\sigma_0' \left(\frac{ \bar Q_{0}^{2}}{M^{2}}\right)^{n} \, \sum_{m = 0}^{\infty} \ \int\limits_{x_F}^{1}dz \,a_{m}^{(2n)j} \wp(x_{F}/z) \, \left(\frac{\bar{\alpha}_{s}Y}{t}\right)^{\frac{m}{2}} I_{m} \left(2\sqrt{\bar{\alpha}_{sS}Y t}\right),
\end{equation}
where $I_m$ is the modified Bessel function of the first kind. The first few coefficients $a_{m}^{(2n)j}$ of the above expansion are presented in Appendix \ref{coeff_a_ex}. An important property of coefficients $a_{m}^{(2n)j}$ is the dependence on $Y$ as a function of the twist: $a_{m}^{(2n)j} \sim \exp(-2(n-1)Y\bar \alpha_{s})$, that leads to a general conclusion that the higher twist contributions to the LO BFKL amplitudes decrease exponentially with the rapidity $Y$.

\section{Results for the helicity structure functions}
\label{results_unint}

In this and the next sections we present results of explicit calculations of the forward Drell--Yan structure functions in $pp$ collisions assuming the LHC energy $\sqrt{S} = 14$~TeV, and the Drell--Yan 
pair $x_F = 0.05$. The calculations of the structure functions are carried out in the Gottfried--Jackson frame. Some earlier results for GBW model were presented in \cite{Stebel:2016pbc}.

There are several possible definitions of the Drell--Yan structure functions which are used for data presentation. The dimensionless structure functions $A_i$ \cite{CSframe} that can be directly related to coefficients of the lepton angular distribution in the DY pair center of mass frame:
\begin{equation}
A_0 = \frac{W_L }{W_{\textrm{tot}}}, \ \ \ A_1 = \frac{W_{LT} }{W_{\textrm{tot}}} , \ \ \ A_2 = \frac{2W_{TT} }{W_{\textrm{tot}}}
\end{equation}
are ratios of the structure functions $W_j$ and $W_{\textrm{tot}}=W_T+W_L/2$. To assess higher twists effects we calculate also exact (sum of all twists) structure functions by evaluation of integral (\ref{Wphihat}) numerically.

\begin{figure}
\begin{center}
\includegraphics[width=.49\textwidth]{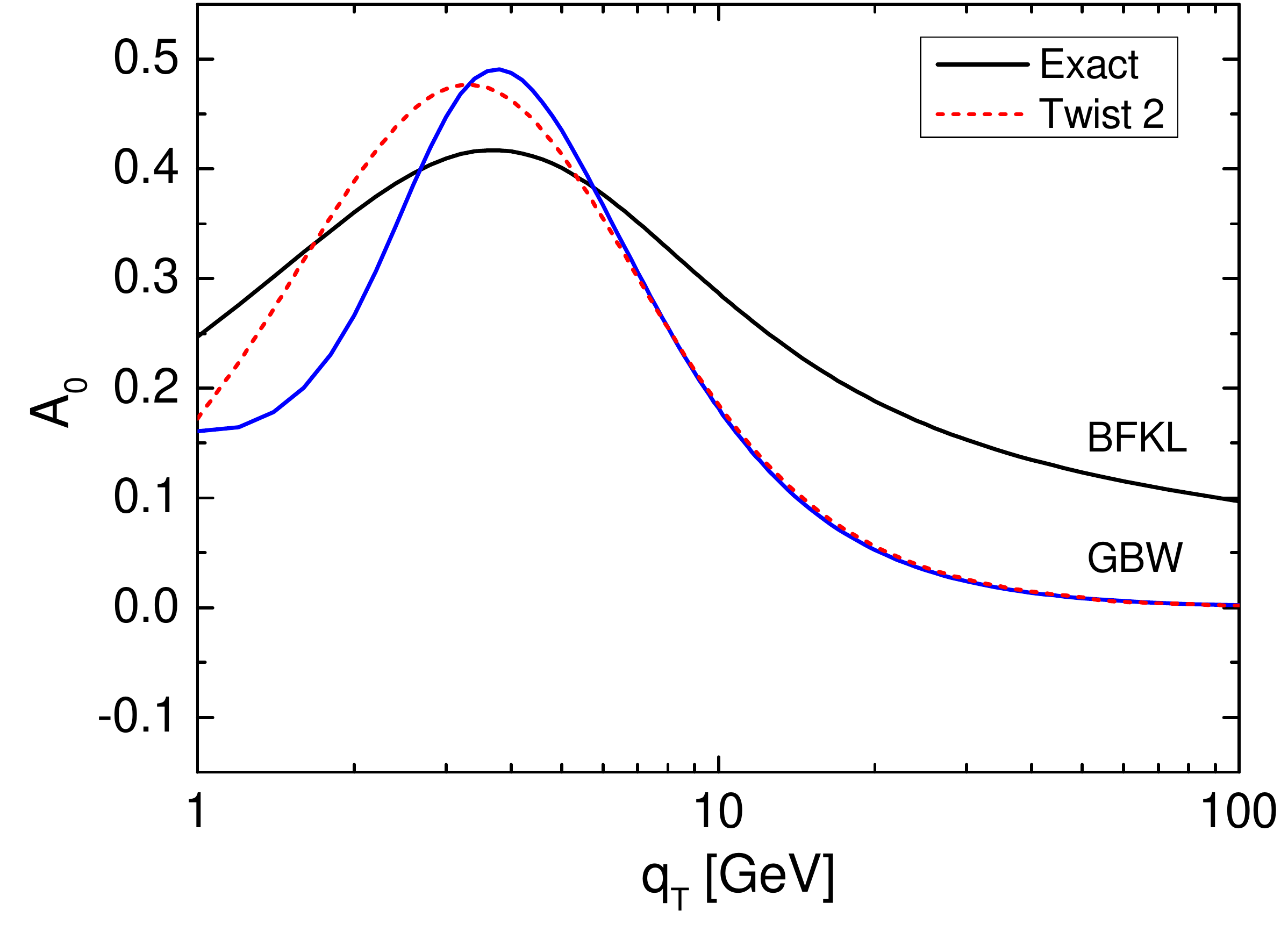}
\includegraphics[width=.49\textwidth]{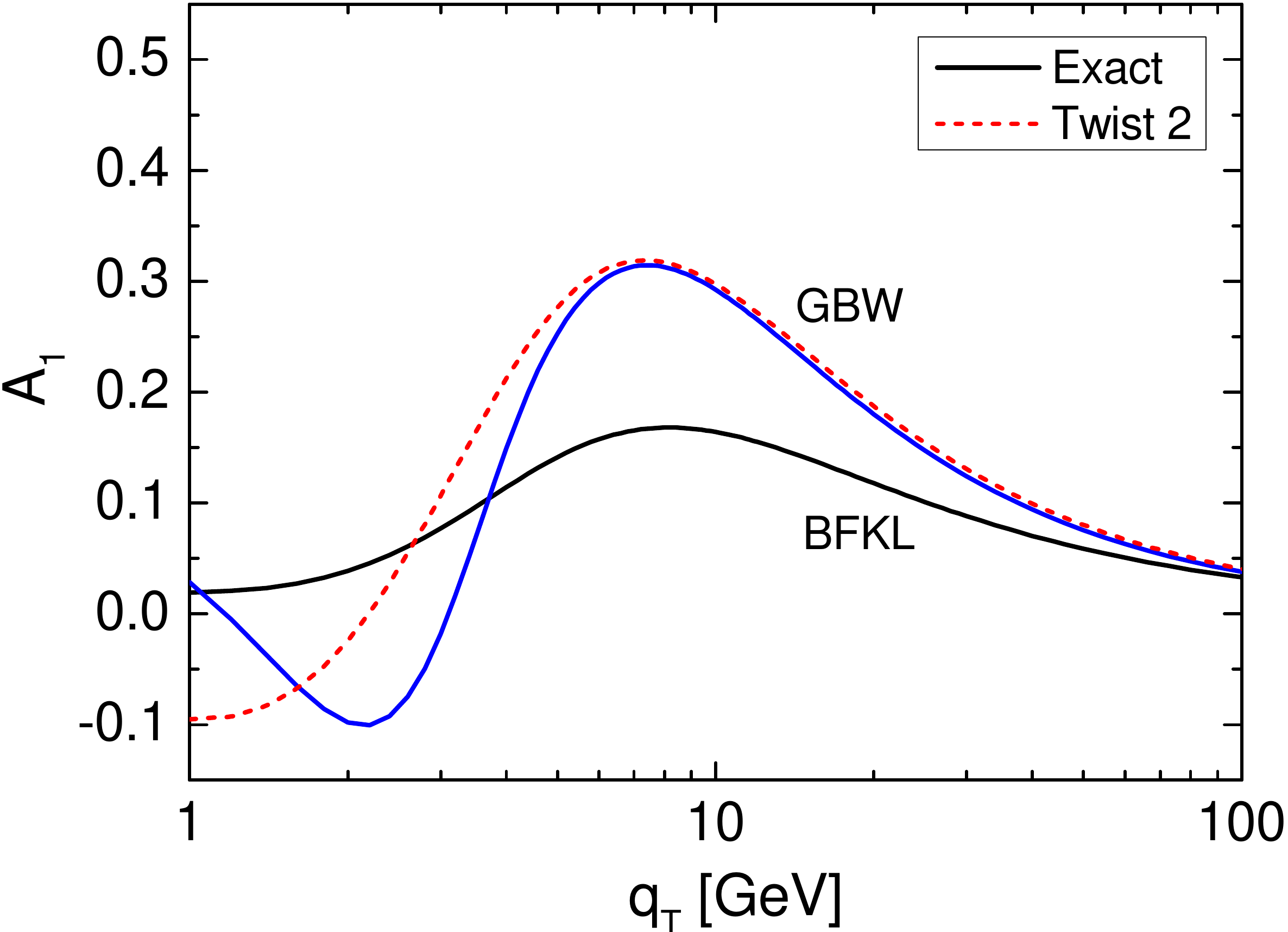} \\
\includegraphics[width=.49\textwidth]{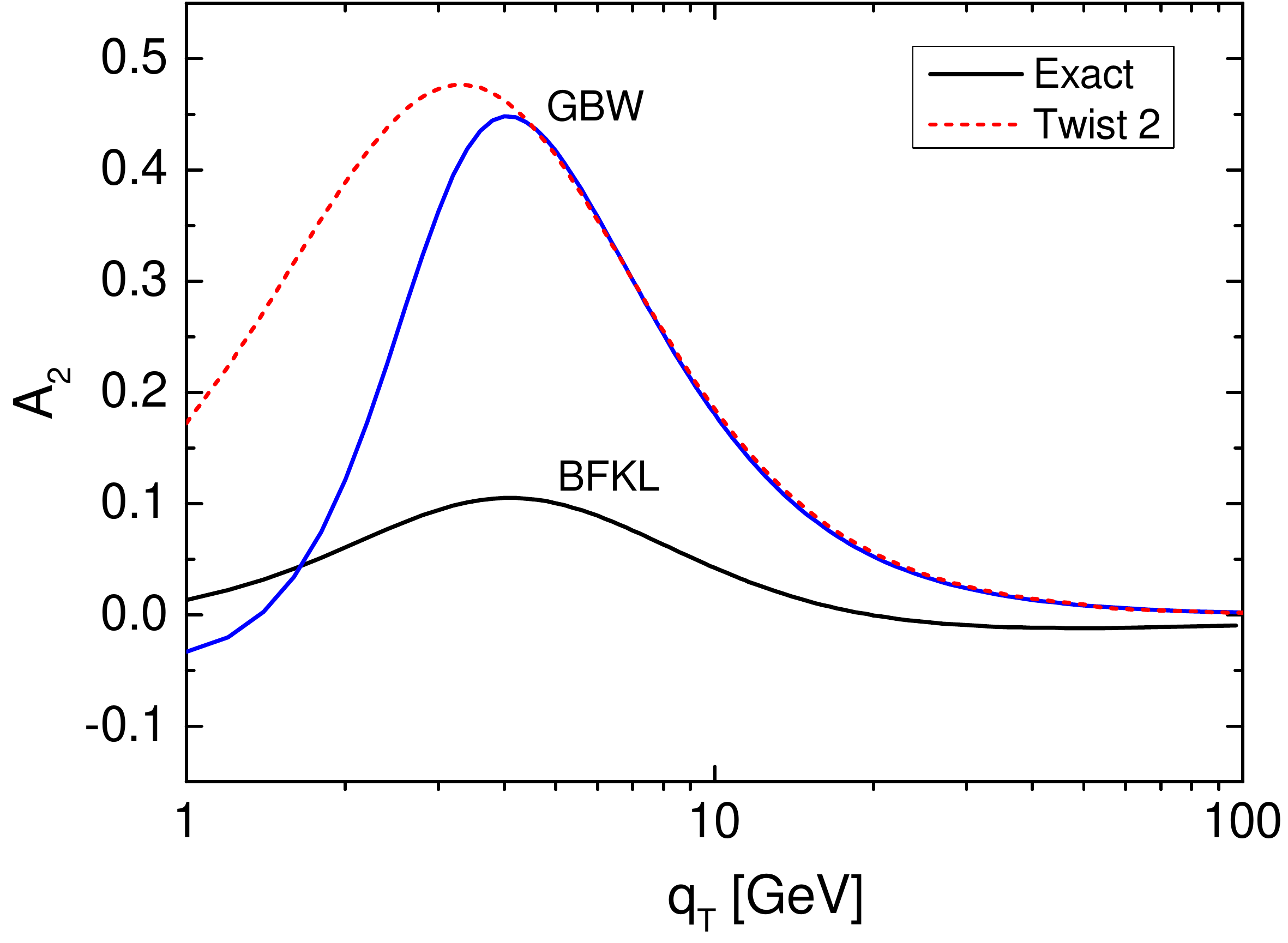} \\
\end{center}
\caption{Coefficiens $A_0$ (top, left), $A_1$ (top, right) and $A_2$ (bottom) as functions of transverse momentum $q_T$ of the intermediate boson. All plots are for $M^2=20$~GeV$^2$.
}
\label{A0_A2_comparison}
\end{figure}

In Fig.\ \ref{A0_A2_comparison} we show a comparison between the exact BFKL results, the exact GBW and the twist~2 GBW components for $A_0$, $A_1$ and $A_2$. We do not show the separated BFKL twist~2 results, as they cannot be distinguished from the exact results, similarly to the case of the DIS analysis presented in \cite{MoSad}. The suppression of the higher twist contribution in the BFKL approach is explicitly illustrated in Fig.\ \ref{coefficientsW_ratios}. It clearly follows from Fig.\ \ref{A0_A2_comparison} that there is a substantial difference between the exact GBW and BFKL predictions. The difference between the models predictions follows mostly from the fact that the BFKL approach takes into account the transverse momentum parton distribution which is strongly limited in the GBW case (exponentially dumped). On the other hand the GBW model predicts sizable contributions from the higher twist terms in contrast to the BFKL expectations dominated by the leading twist term. Therefore, both models may serve as good benchmarks for the competition between the transverse momentum distributions and higher twist effects.

\begin{figure}
\centering
\includegraphics[width=.49\textwidth]{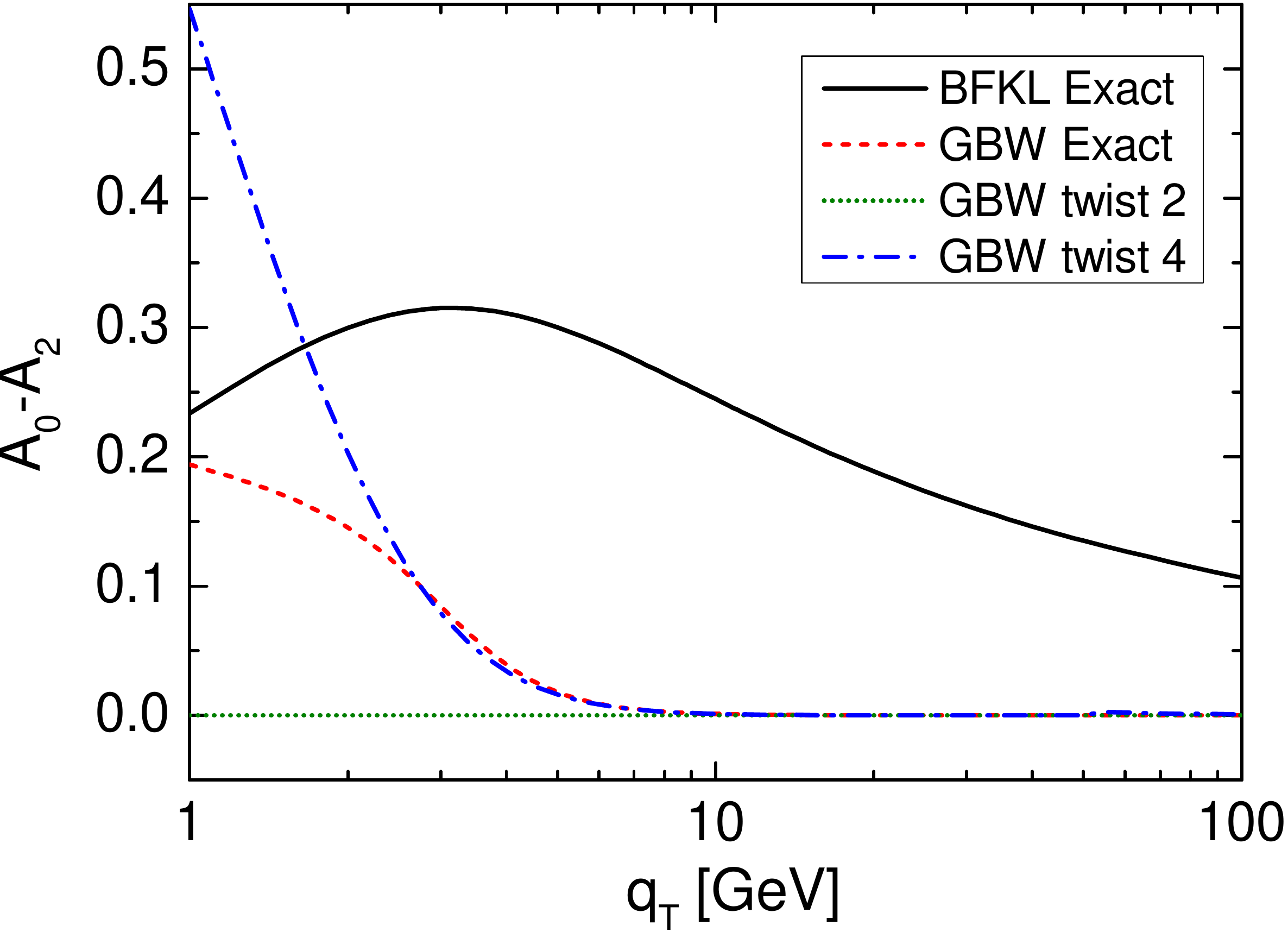}
\includegraphics[width=.49\textwidth]{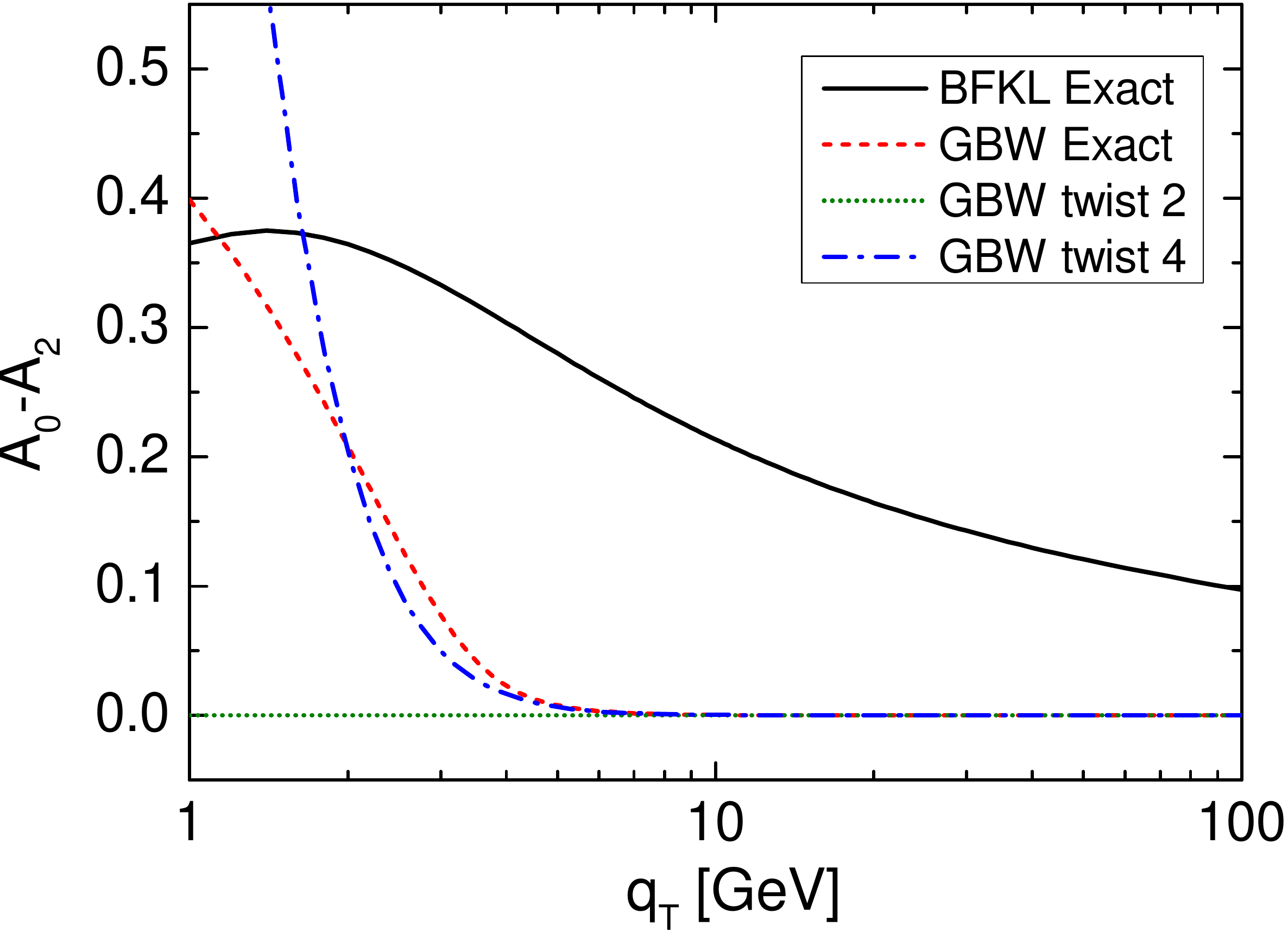}
\caption{
The Lam--Tung observable $A_{LT}$ as a function of $q_T$, at $M^2 = 20$ GeV$^2$ (left) and $M^2 = 5$ GeV$^2$ (right) assuming the GBW and BFKL scattering amplitudes. For the GBW we show the twist~2 contribution (vanishing), the twist~4 contribution and the results obtained summing all twists (``GBW Exact''). For the BFKL cross section the twist~2 component does not differ from the all-twist-sum (``BFKL Exact'').
}
\label{BFKL_GBW_A_lam}
\end{figure}

An important observable in the analysis of subtle QCD effects beyond the leading twist collinear approximation, is the Lam--Tung observable $A_{LT} = A_0 - A_2$. In the collinear QCD framework $A_{LT}=0$ up to the next-to-next-to leading order. Hence $A_{LT}$ is considered to be a good probe of the higher twist effects and parton transverse momentum effects which do not compete here with the leading twist collinear contributions. 
In the left pannel of Fig.\ \ref{BFKL_GBW_A_lam} the Lam--Tung observable $A_{LT} = A_0 - A_2$ is shown as a function of the transverse momentum of the lepton pair at $M^2=20$~GeV$^2$. In the GBW model the twist~2 contribution is consistent with
$A_{LT} = 0$, that is for the leading twist GBW the Lam--Tung relation is preserved. This follows explicitly from the analytic expressions for the twist expansion of the forward DY structure functions. It is expected as the color dipole description of the forward DY process is based on the partonic diagrams with the topology of the NLO contribution, and the GBW model leads to an almost collinear gluon distribution. At $q_T \lesssim 5$~GeV the deviations of $A_{LT}$ from zero appear in the GBW model, that are driven by the twist~4 term between
$q_T \simeq 3$~GeV and $q_T \simeq 5$~GeV. Below $q_T \lesssim 3$~GeV twist~4 contribution is not sufficient and even the higher twist contributions become relevant. Within the exact GBW model, the Lam--Tung relation is broken at the level of 0.1 -- 0.2 for $q_T < 3$~GeV due to the higher twist effects. Hence, within the GBW model of the color dipole cross section, the higher twist effects are clearly visible at lower values of the lepton pair transverse momentum. 

The pattern following from the BFKL scattering amplitudes is different. Significant violation of the Lam--Tung relation occurs at all of the probed $q_T$ range. Recall that the BFKL amplitudes lead to negligible higher twist contributions. Hence, already the leading twist contribution of the BFKL model strongly violates the Lam--Tung relation. This breaking effect is a consequence of the wide transverse momentum distribution of the virtual gluons coming from the BFKL evolution (the more detailed study of the transverse momentum effects in the Lam--Tung relation breaking, taking into account also the $g^*g^*$ channel, was performed in Ref.\ \cite{MoSadSteLT}). In fact, the gluon transverse momentum effects in the BFKL scattering amplitude lead to stronger breaking of the Lam--Tung relation than the higher twist effects in the GBW model down to $q_T = 2$~GeV. At lower $q_T$, however, the higher twist effects from the GBW and the gluon transverse momentum effects from the BFKL are similar. 

In the right panel of Fig.\ \ref{BFKL_GBW_A_lam} we show the Lam--Tung observable in both models for a very low mass of the DY pair, $M^2 = 5$~GeV$^2$. At this mass we find a similar pattern to the case of $M^2 = 20$~GeV$^2$, with slightly enhanced both the higher twist and the gluon transverse momentum effects. It follows from Fig.\ \ref{BFKL_GBW_A_lam} that for the Lam--Tung observable at low masses, lowering the mass does not lead to relative amplification of the higher twist corrections w.r.t.\ the parton transverse momentum effects. Hence, even at lower DY pair masses, even at low $q_T$, and very large energies and in the forward kinematics, the effects of parton transverse momentum and of the higher twists are expected to have a similar contribution to the Lam--Tung relation breaking. At higher masses and at larger $q_T$ the higher twist terms become small, but the possible effects of parton $k_T$ may still stay sizable, see e.g.\ Ref.\ \cite{MoSadSteLT}). 
These results indicate that the observation of the higher twist contributions in $A_{LT}$ requires a good control of the parton transverse momentum effects. 

\section{Twist expansion of the integrated helicity structure functions}
\label{twist_exp_int}

It may be useful to study experimentally the forward Drell--Yan structure functions integrated over $q_T$. They are defined in the following way,
\begin{equation}
\tilde{W}_j = \frac{1}{2\pi M^2} \int W_j ~ d^2q_T
\end{equation}
and their Mellin representation takes the form:
\begin{eqnarray}
\tilde{W}_{j} & = &\int_{\cal C} \frac{ds}{2\pi i}\ \int_{x_F}^1 dz \ \wp(x_F/z) f_j(z) \left ( \frac{z^2 \bar Q_0^2}{4 M^2(1-z)} \right) ^{-s} \tilde{\sigma}(s,Y) H_j(-s),
\end{eqnarray}
where:
\begin{equation}
f_T(z)=\frac{1+(1-z)^2}{z^2}, \ \ f_L(z)=f_{TT}(z)=\frac{1-z}{z^2}, \ \ f_{LT}=\frac{(2-z)\sqrt{1-z} }{z^2}.
\end{equation}
The rapidity evolution length was defined in (\ref{Y_def}) and for $x_g$ we use threshold ($q_T \rightarrow 0$) value of (\ref{xg_def}), $x_g=M^2/(Sx_F)$ . The expressions for the Mellin transformed impact factor $H_i(s)$ were derived in \cite{GBLS,MoSadSte} and are listed in Appendix \ref{impact_facts_int}. The formulae for the twist decomposition of the $q_T$-integrated forward DY structure functions assuming the GBW dipole cross section were given in \cite{MoSadSte}. Below we derive the twist expansions for the $q_T$-integrated DY structure functions with the BFKL exchange using the procedure described in Sec.\ \ref{twExp}.
Hence we insert the explicit Mellin representation of the BFKL cross section into the above formula for the integrated structure functions, 
\begin{eqnarray}
\tilde{W}_{j} & = & -\sigma_{0}' \int_{x_F}^1 dz \ \wp(x_F/z) f_j(z) \int_{\cal C} \frac{ds}{2\pi i}\ e^{s \tilde{t}} \ \Gamma(s) e^{\bar{\alpha}_s \chi(s) Y}
\left( \frac{z^2}{1-z} \right)^{-s} H_j(-s),
\end{eqnarray}
where $\tilde{t}=\textrm{ln} ( 4M^2/\bar Q_0^2 )$.
Following the steps described in the previous section one gets the expression for the twist-$2n$ component of the structure functions:
\begin{eqnarray}
\label{Wi2n}
\tilde{W}_{j} ^{(2n)} &=& -\sigma_{0}' e^{-n\tilde{t}} \int_{x_F}^1 dz \ f_j(z) \left( \frac{z^2}{1-z} \right)^n \ \wp(x_F/z) \nonumber \\
& &\times \ \int_0^{2\pi} d \theta \ \tilde{h}_j^{(2n)}(\epsilon e^{i\theta},z,Y) 
\left( \frac{z^2}{1-z} \right)^{-\epsilon \ \exp i \theta}
\exp \left( \epsilon e^{i \theta} \ \tilde{t} + \bar{\alpha}_s Y \frac{1}{ \epsilon e^{i \theta} } \right),
\end{eqnarray}
where the coefficients $\tilde{h}_j^{(2n)}$ read
\begin{equation}
\tilde{h}_j^{(2n)}(\epsilon e^{i\theta},z,Y) = \epsilon e^{i\theta} H_{j}(n-\epsilon e^{i\theta},z) \Gamma(-n+\epsilon e^{i\theta}) e^{\bar{\alpha}_{s} Y \chi^{(n)}_{reg}},
\end{equation}
cf.\ the analogous expression (\ref{hn2}) for the coefficients $h_j^{(n)}$ in the $q_T$-integrated case discussed in section \ref{twExp}. For $j=L,TT,LT$, after the integration over $\theta$, the leading twist terms are given by,
\begin{eqnarray}
\tilde{W}_{j} ^{(2)} &=& -\sigma_{0}' \left( \frac{\bar Q_0^2}{4M^2} \right) \int_{x_F}^1 dz \ f_j(z) \frac{z^2}{1-z} \ \wp(x_F/z) \nonumber \\
& &\times \ \sum_{m=0}^{\infty} \tilde{a}_m^{(2)j}\left( \frac{\bar{\alpha}_s Y}{ \textrm{ln} ( 4M^2/\bar Q_0^2 ) } \right)^{\frac{m}{2}} I_{|m|} \left( 2\sqrt{\bar{\alpha}_s Y \textrm{ln} \frac{4M^2}{\bar Q_0^2} } \right)
\end{eqnarray}
where $\tilde{a}_m^{(2n)j}$ are the expansion coefficients of the functions
\begin{equation}
\tilde{h}_j^{(2n)}(\epsilon e^{i\theta},z,Y) \left( \frac{z^2}{1-z} \right)^{-\epsilon  \exp i \theta} = \sum_{m=0}^{\infty} \tilde{a}_m^{(2n)i}\ \left( \epsilon e^{i \theta} \right)^m ,
\label{series_hbar}
\end{equation}
where the dependence of $\tilde{a}_m^{(2n)j}$ on the variables $z$ and $Y$ was suppressed.
The first two leading coefficients $\tilde{a}_m^{(2)j}$ for $i=L,TT,LT$ read: 
\begin{eqnarray}
\tilde{a}_0^{(2)L} &=& -\frac{4}{3}, \ \ \ \tilde{a}_1^{(2)L} = -\frac{4}{3} \left( -2 +2 \gamma_E + \textrm{ln}\frac{1-z}{z^2} + \psi(5/2) \right), \\
\tilde{a}_0^{(2)TT} &=& -\frac{2}{3}, \ \ \ \tilde{a}_1^{(2)TT} = \frac{2}{3} \left( -3 + \gamma_E - \textrm{ln}\frac{1-z}{z^2} + \textrm{ln}(64) + 2\psi(5/2) \right), \\
\tilde{a}_0^{(2)LT} &=& 0, \ \ \ \tilde{a}_1^{(2)LT} = 0.5236.
\end{eqnarray}
Note that for the most leading coefficients $\tilde{a}_0^{(2)j}$ the Lam--Tung relation is preserved: $\tilde{a}_0^{(2)L}=2\tilde{a}_0^{(2)TT}$. This follows from the fact that the most leading coefficients in the $\epsilon$ expansion correspond to the double logarithmic approximation of the BFKL exchange which coincides with the collinear approximation results. As it is expected, the Lam--Tung relation is broken by the non-leading coefficients $\tilde{a}_m ^{(2)j}$, $m>0$, that correspond to the BFKL effects beyond the double logarithmic limit. Also interesting is to note that the leading twist 2 coefficient $\tilde{a}_0^{(2)LT}=0$ for the structure function $\tilde W_{LT}$. As the analytical expression for $H_{LT}$ is not known, the expansion coefficient $\tilde{a}_1^{(2)LT}$ and the coefficients $\tilde{a}_m^{(2)LT}$, $m>1$ (not listed) were obtained only numerically.

A more sophisticated procedure is necessary to obtain the twist~2 component of $\tilde W_{T}$ and the twist components of all the structure functions beyond twist~2, because a logarithmic divergence of the form 
$\ln (1-z)$ occurs in the integrals corresponding to the twist components at $z \to 1$. A treatment of such apparent singularities was developed in \cite{GBLS} and we follow that procedure --- see Appendix \ref{twist_T} for the details.

\begin{figure}
\centering
\includegraphics[width=.6\textwidth]{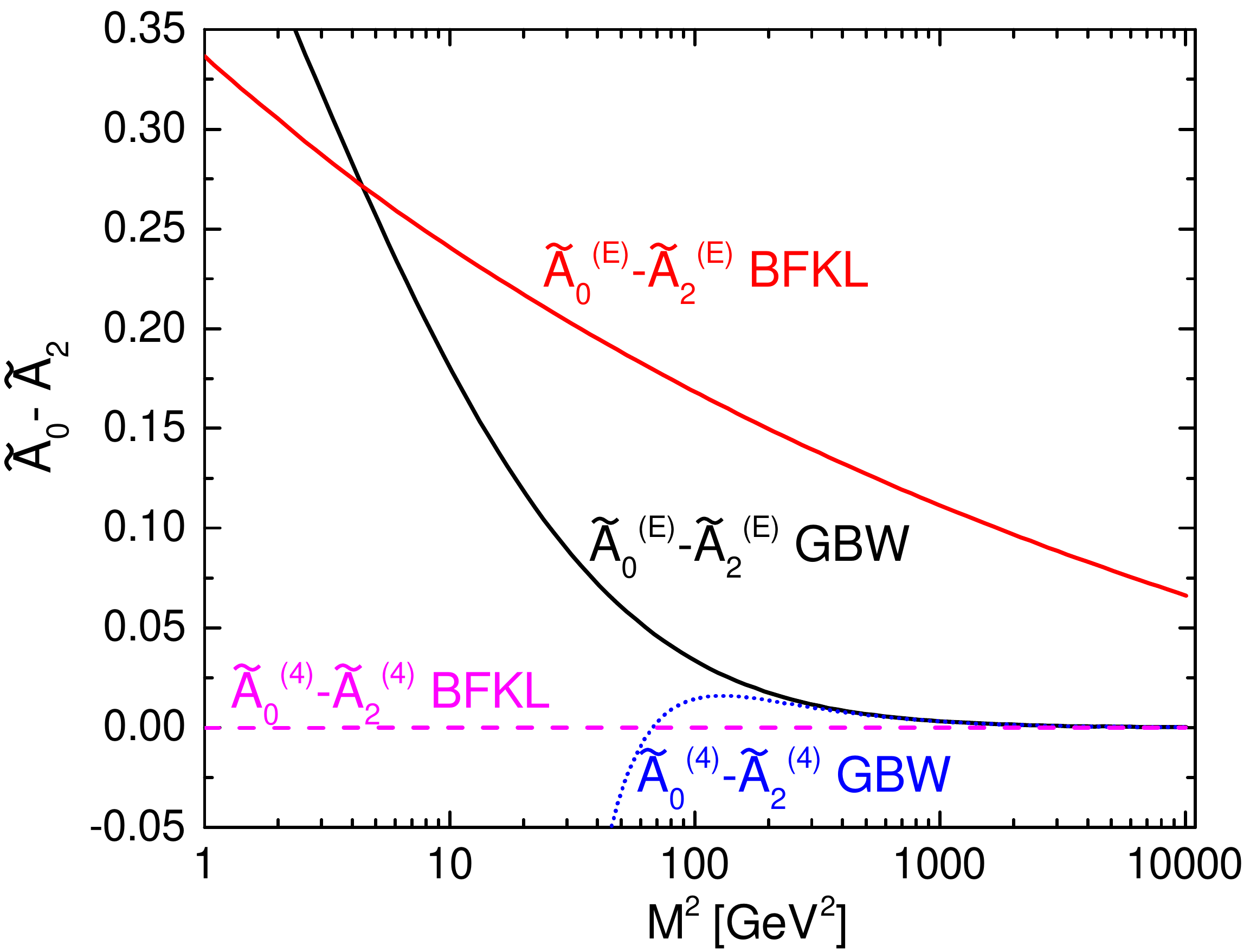}
\caption{The Lam--Tung combination $\tilde A_{LT} =\tilde{A}_0- \tilde{A}_2$ for the exact (sum of all twists) values and the twist~4 component in the GBW and BFKL models.}
\label{LamTung_and_A}
\end{figure}

\section{Results for the integrated helicity structure functions}
\label{results_int}

The results in this section are presented using the dimensionless coefficients $\tilde{A}_i$:
\begin{equation}
\tilde{A}_0 = \frac{\tilde{W}_L }{\tilde{W}_{\textrm{tot}}}, \ \ \ \tilde{A}_1 = \frac{\tilde{W}_{LT} }{\tilde{W}_{\textrm{tot}}} , \ \ \ \tilde{A}_2 = \frac{2\tilde{W}_{TT} }{\tilde{W}_{\textrm{tot}}}.
\end{equation}
where $\tilde{W}_{\textrm{tot}}=\tilde{W}_T+\tilde{W}_L/2$. Additionally, we introduce coefficients $\lambda_1$, 
$\lambda_2$ and $\lambda_3$ which are invariant with respect to rotations in the $X - Z$ plane in the lepton center of mass frame \cite{Palestini:2010xu}. They read
\begin{equation}
\lambda_{1} = \frac{\lambda_{\theta}+3\lambda_{\phi} }{1-\lambda_{\phi}} , \ \ \ \lambda_{2} = \frac{ 1+(\lambda_{\theta}-\lambda_{\phi})/4 }{\sqrt{(\lambda_{\theta}-\lambda_{\phi})^2+ 4 \lambda_{\theta \phi}^2}} , \ \ \ \lambda_{3} = 1-\lambda_{\theta}-4\lambda_{\phi},
\end{equation}
where
\begin{equation}
\lambda_{\theta} = \frac{\tilde{W}_T-\tilde{W}_L }{\tilde{W}_T+\tilde{W}_L} , \ \ \ \lambda_{\phi} = \frac{\tilde{W}_{TT} }{\tilde{W}_T+\tilde{W}_L} , \ \ \ \lambda_{\theta \phi} = \frac{\tilde{W}_{LT} }{\tilde{W}_T+\tilde{W}_L}.
\end{equation}

In Fig.\ \ref{LamTung_and_A} the Lam--Tung observable $\tilde A_{LT} = \tilde{A}_0- \tilde{A}_2$ for the $q_T$-integrated distributions is presented for the exact values of the GBW and BFKL models $\tilde{A}^{(E)}_{LT}$ and for the twist~4 components of the models $\tilde{A}^{(4)}_{LT}$. Recall that in GBW model the twist~4 component provides the leading contribution to the Lam--Tung observable. It follows from the figure that in the GBW model the higher twist contributions in $\tilde A_{LT}$ become visible below $M^2=100$ GeV$^2$. However, the BFKL model predicts sizable violation of the Lam--Tung relation in the integrated DY structure functions in the whole plotted range of the masses (at the leading twist). The origin of the violation may be traced back to the strong parton transverse momentum effects in the BFKL approach. These effects are stronger than the GBW higher twist effects down to $M^2=4$ GeV$^2$. Only below that threshold the higher twists prevail. This supports our previous conclusion that disentanglement of the higher twist effects from the parton transverse momentum effects requires a careful analysis and acquiring a good understanding of the parton $k_T$ distribution. In BFKL twist 4 is negligible so $\tilde{A}^{(4)}_{LT}$ is very close to zero.

\begin{figure}
\centering
\includegraphics[width=.6\textwidth]{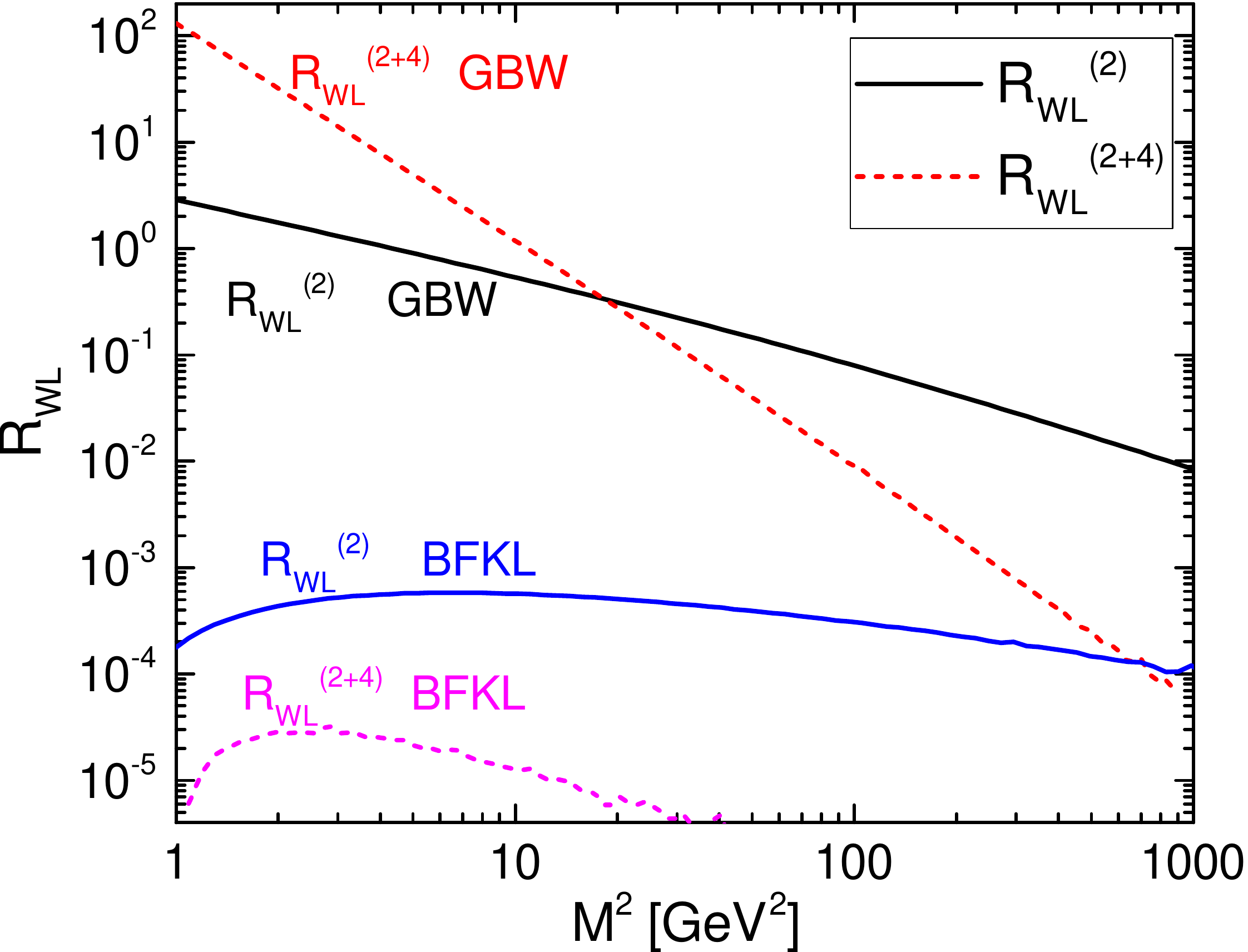}
\caption{Ratios $R_{WL}^{(2)}=\tilde{W}_L^{(2)}/\tilde{W}_L^{(E)}-1$ and $R_{WL}^{(2+4)}=(\tilde{W}_L^{(2)}+\tilde{W}_L^{(4)})/\tilde{W}_L^{(E)}-1$ in the GBW and BFKL models. By $\tilde{W}_L^{(E)}$ we understand the sum of all twist components.}
\label{coefficientsW_ratios}
\end{figure}

\begin{figure}
\centering
\includegraphics[width=.6\textwidth]{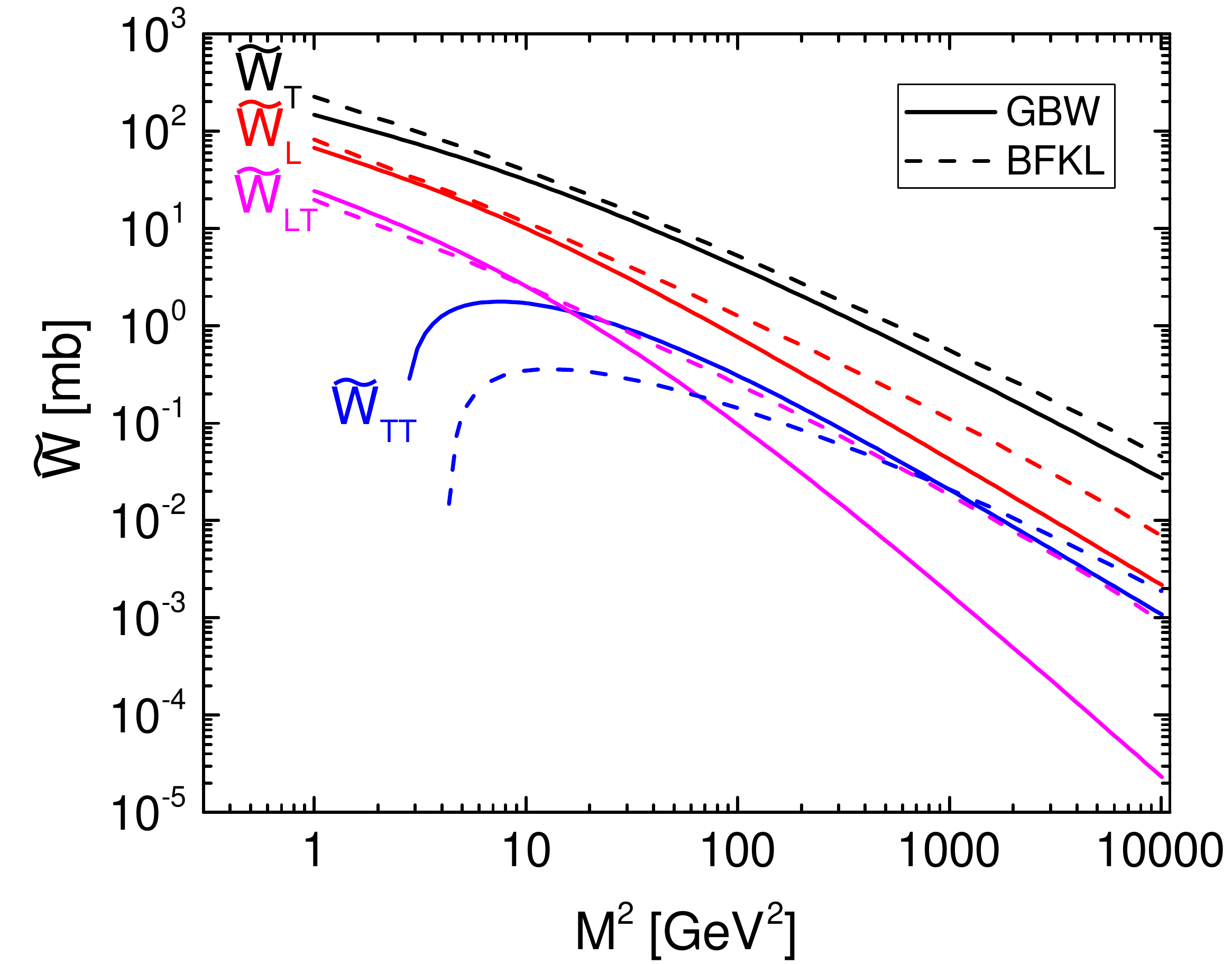}
\caption{$M^2$-dependence of the $q_T$-integrated DY structure functions $\tilde W_j$ obtained from the GBW and BFKL models. }
\label{coefficientsW_exact}
\end{figure}

In Fig.\ \ref{coefficientsW_ratios} we illustrate the higher twist contributions to a selected structure function $\tilde W_L$ using ratios: $R_{WL}^{(2)}=\tilde{W}_L^{(2)}/\tilde{W}_L^{(E)}-1$ and $R_{WL}^{(2+4)}=(\tilde{W}_L^{(2)}+\tilde{W}_L^{(4)})/\tilde{W}_L^{(E)}-1$. $R_{WL}^{(2)}$ ($R_{WL}^{(4)}$) is the relative negative contribution of the twist~$n$ components with $n>2$ ($n>4)$. The figure shows that the BFKL result is dominated by the leading twist component, and the higher twist components enter at the level of $10^{-3}$ of the dominant twist~2 term for the whole range of $M^2 > 1$ GeV$^2$.
In the GBW model the twist~4 correction to $\tilde W_L$ becomes relevant below $M^2 \simeq 100$~GeV$^2$, and below $M^2 \simeq 30$~GeV$^2$ all the twist components should be taken into account.

\begin{figure}
\centering
\includegraphics[width=.455\textwidth]{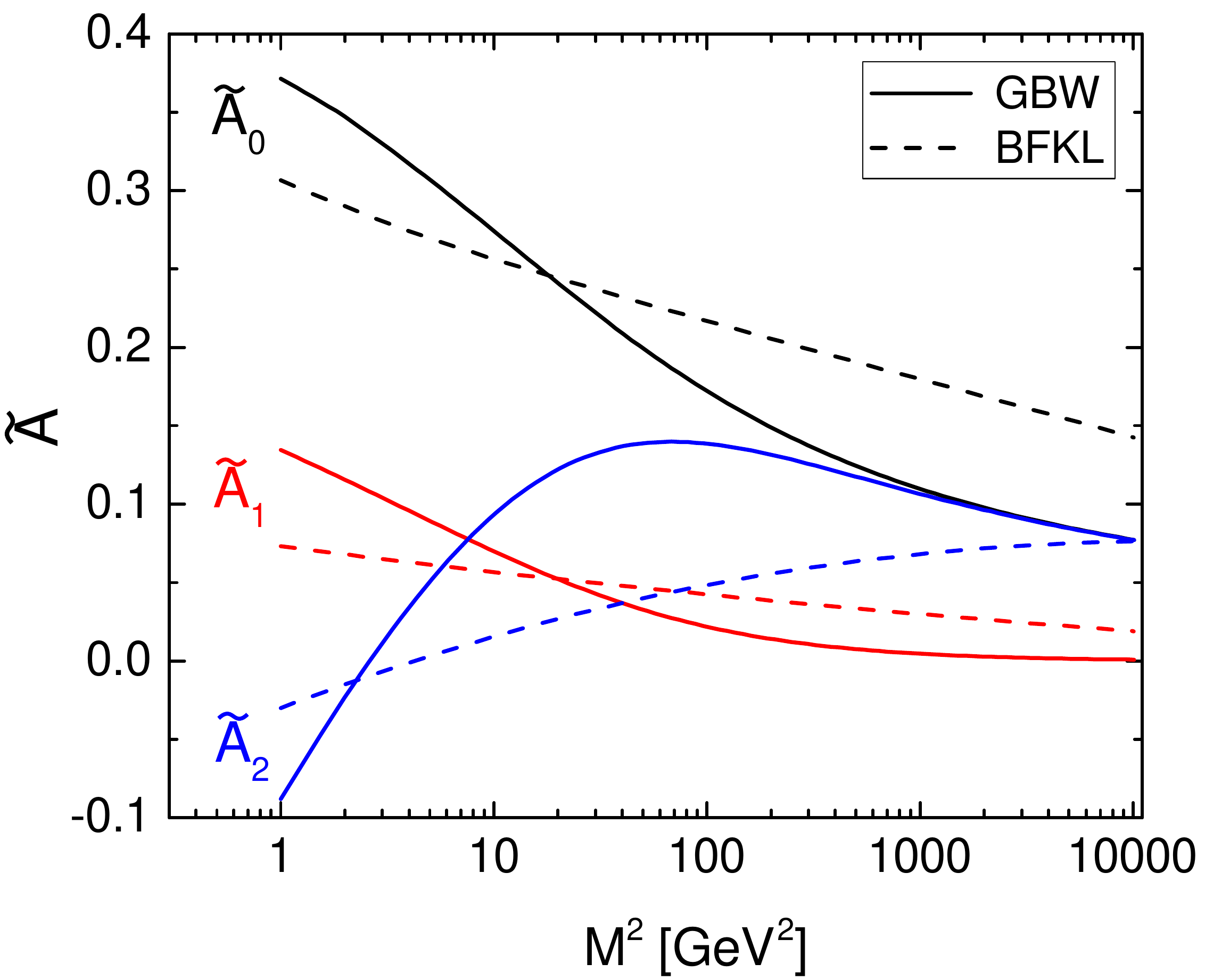}
\includegraphics[width=.49\textwidth]{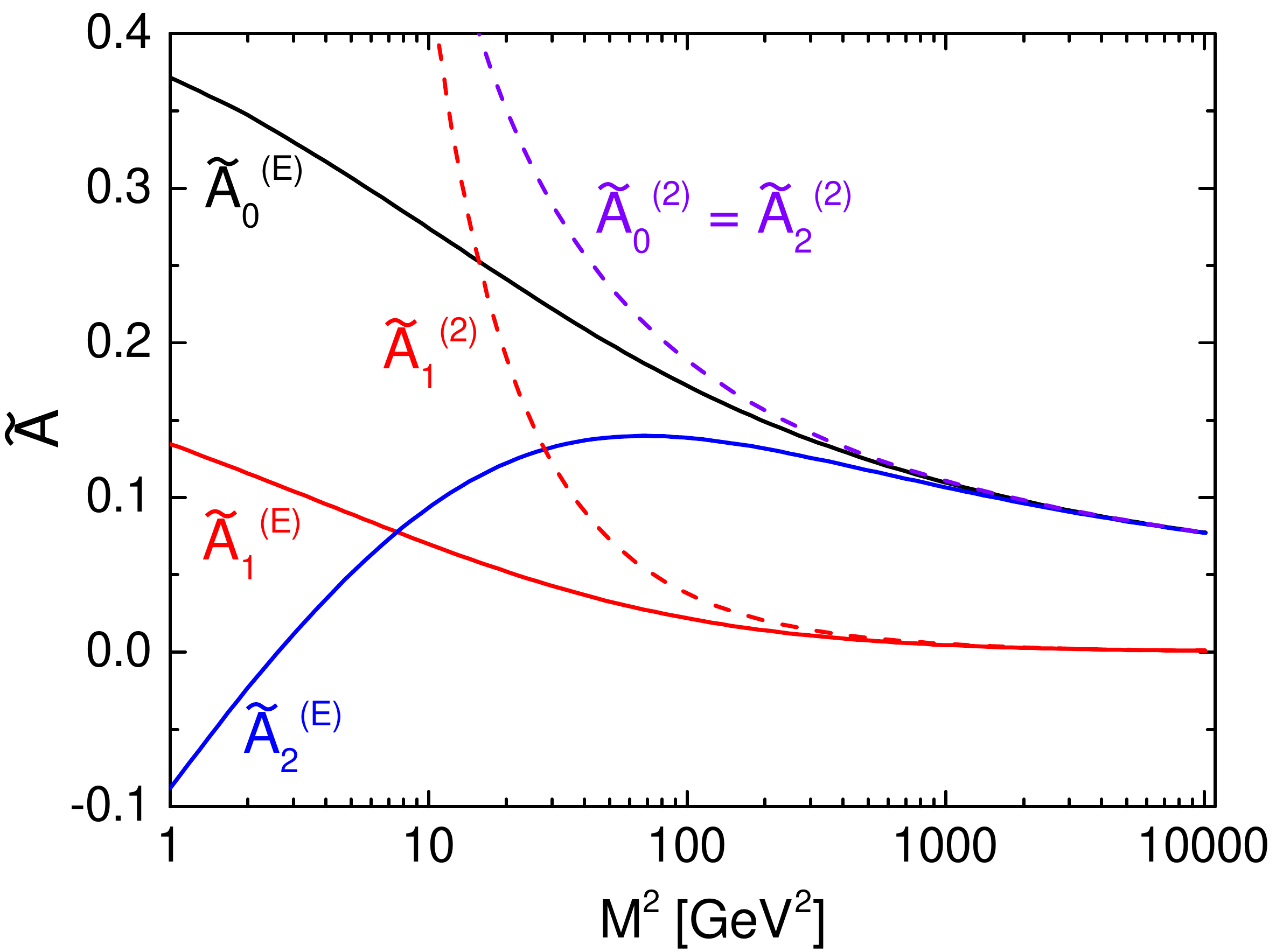}
\caption{$M^2$-dependence of the angular coefficients $\tilde A_i$. Left: comparison of the
exact values obtained in the BFKL and GBW models. 
Right: results of the GBW model --- the twist~2 component compared to the exact results. 
Note that in the GBW model the twist~2 components of $\tilde A_0$ and $\tilde A_2$ coincide. }
\label{fig:Acoefficients}
\end{figure}

\begin{figure}
\centering
\includegraphics[height=60mm, width=80mm]{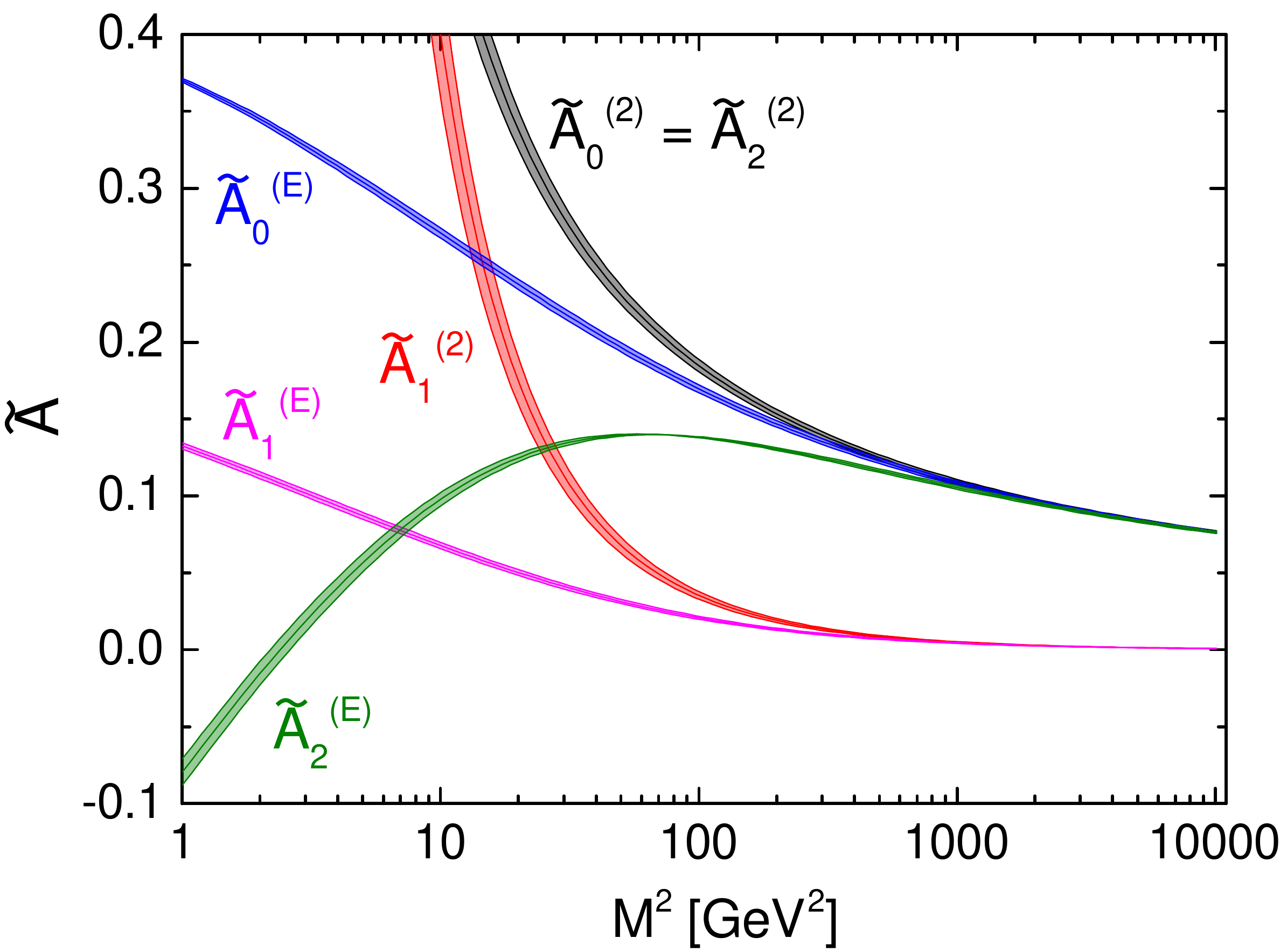}
\caption{Sensitivity of the GBW model predictions to the choice of the $x_0$ model parameter: the angular coefficients $\tilde A_i$ as functions of $M^2$  --- the twist~2 components compared to the exact results. The bands reflect the spread due to variation between the original GBW parameter $x_0$ and $\hat x_0$. }
\label{A_GBW_error_bars}
\end{figure}

Fig.\ \ref{coefficientsW_exact} shows comparison between the predictions of the GBW and BFKL models for $\tilde{W}_i$ structure functions (the exact values). For $\tilde W_{L}$, $\tilde{W}_T$, and $\tilde W_{LT}$ the largest differences between the predictions appear at higher values of $M^2$. In this kinematical range the higher twist contributions are suppressed and the shape of the curves is determined by the transverse momentum effects. However, the $\tilde W_{TT}$ structure function exhibits the opposite behavior. The largest difference between the BFKL and GBW models appears at the lower values of $M^2$. One concludes that $\tilde W_{TT}$ is the particularly sensitive structure function to the higher twist effects.

The same pattern is found in the dimensionless integrated structure functions $\tilde A_i$. On the left hand side of Fig.\ \ref{fig:Acoefficients} we show a comparison between predictions of $\tilde A_i$ obtained with the GBW and BFKL models. Comparing this plot with the right panel of Fig.\ \ref{fig:Acoefficients} one concludes that the higher twist contribution in the GBW model starts being visible already for $M^2\lesssim 300$~GeV$^2$. 
Note that the coefficient $\tilde{A}_2$, that is directly related to $\tilde W_{TT}$, is particularly sensitive to the effects beyond the leading twist collinear approximation. At the level of twist~2 the difference between the BFKL model and the GBW model results may be treated as an approximate measure of the gluon $k_T$ effects that are sizable in the BFKL approach and almost negligible in the GBW model. Comparing the higher twist content $\tilde A_2 ^{(E)} - \tilde A_2 ^{(2)}$ in the GBW results (see Fig.\ \ref{fig:Acoefficients}, the right panel) with the spread between the GBW and BFKL predictions for $\tilde A_2$, one concludes that for $M^2\lesssim 30$~GeV$^2$ the higher twist effects that follow from the GBW model are larger than the gluon $k_T$ effects following from the BFKL evolution. Therefore $\tilde A_2$ at low $M^2$ should be particularly useful observable for an experimental discrimination between the models and for constraining the higher twist contributions.

As discussed in Sec.\ \ref{strFunc}, we introduced a modified value of the original $x_0$ parameter $\hat x_0 = 2x_0$ in the model of the GBW dipole cross section. In order to display the sensitivity of the GBW predictions to this parameter variation we compare 
in Fig.\ \ref{A_GBW_error_bars} the twist~2 and the exact estimates for the $q_T$-integrated angular coefficients $A_i$ obtained with the original GBW value $x_0$ and the modified value $\hat x_0$. The resulting theoretical uncertainty bands are found to be rather narrow. 

\begin{figure}
\centering
\includegraphics[width=.49\textwidth]{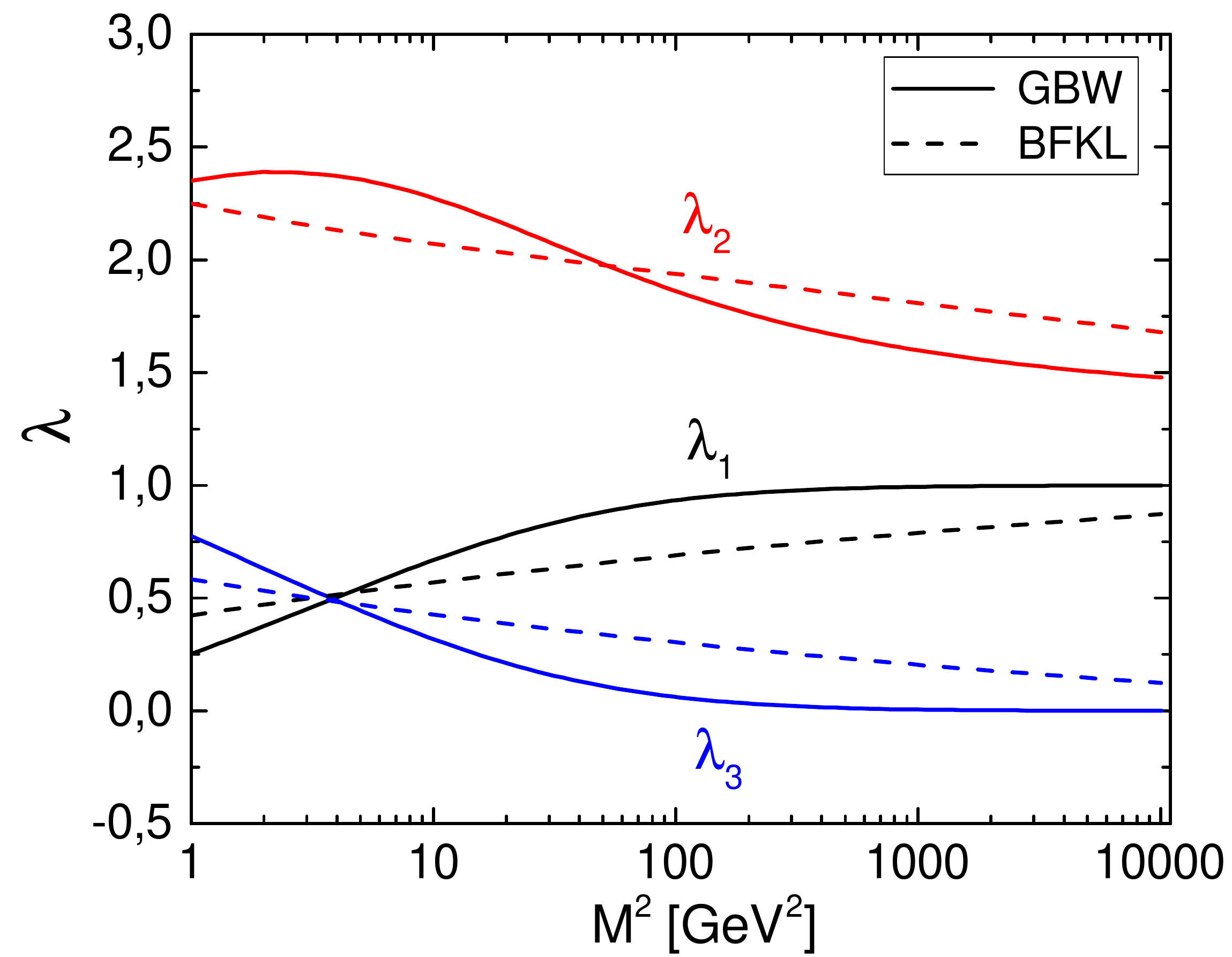}
\includegraphics[width=.49\textwidth]{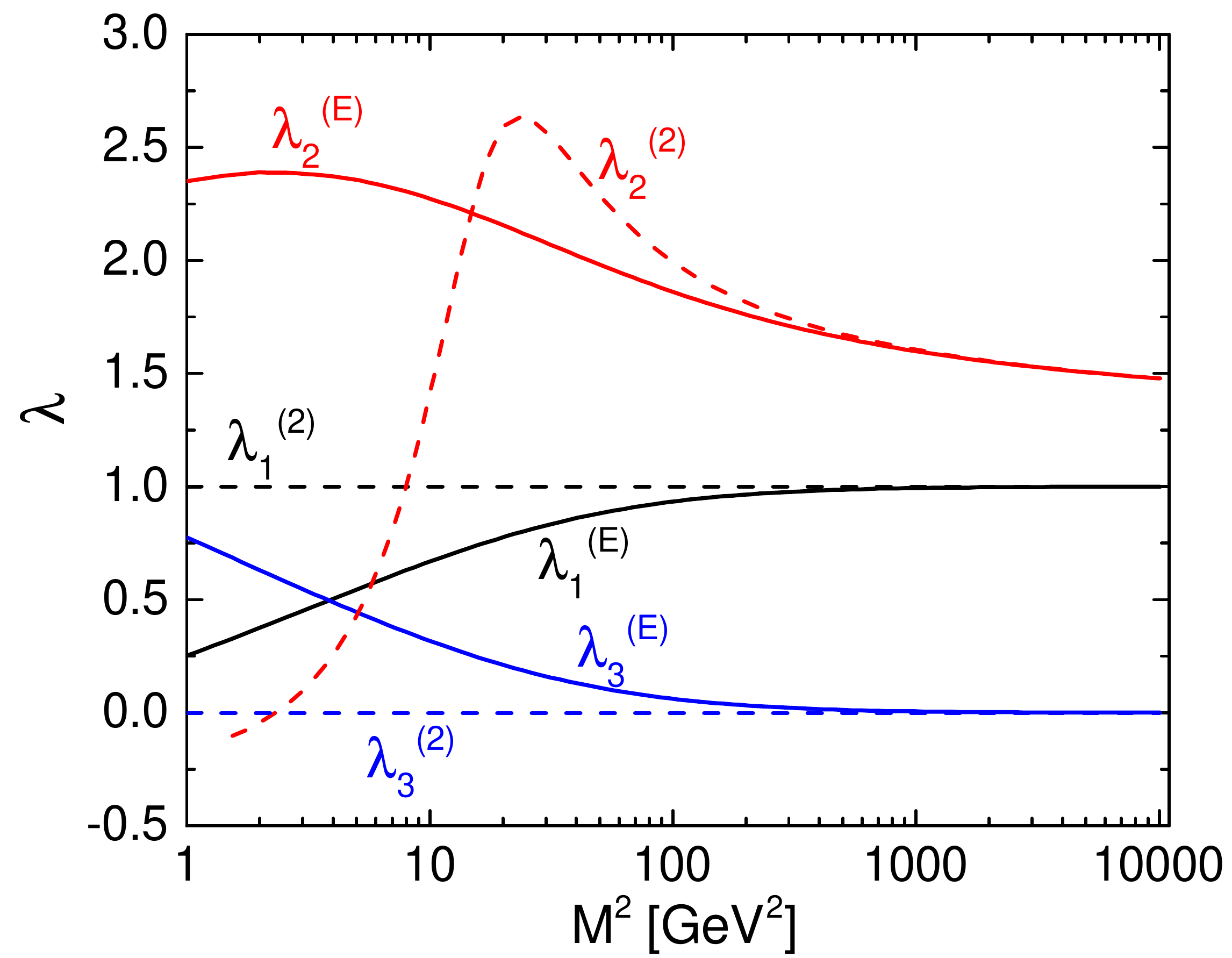}
\caption{
The invariant coefficients $\lambda_i$ as functions of $M^2$. Left: comparison of the
exact values obtained in the BFKL and GBW models. 
Right: results of the GBW model --- the twist~2 component compared to the exact results. 
}
\label{fig:lambdas}
\end{figure}

In Fig.\ \ref{fig:lambdas} we show also the results in terms of invariant coefficients $\lambda_i$, that are valid in the frames with the $Y$-axis transverse to the beam --- DY pair plane \cite{Palestini:2010xu}. Particularly strong higher twist effects in the GBW approach are found in $\lambda_2$ for $M^2\lesssim 100$~GeV$^2$ (see the right panel).
Note also since $\lambda_{\phi} \ll 1$ and $1-\lambda_1 = \lambda_3/(1 -\lambda_{\phi})$ we have $1-\lambda_1 \approx \lambda_3$ which can be seen on the plot. At the leading twist the GBW model yields $\lambda_{1}=1$
and $\lambda_{3}=0$ that implies that in this approximation the Lam--Tung relation $\tilde W_L-2 \tilde W_{TT}=0$ is satisfied, as expected. 

To sum up, the forward Drell--Yan angular coefficients $\tilde{A}_i$ in particular
$\tilde{A}_2$ for $M^2 \lesssim 100$~GeV$^2$ are sensitive to the higher twist effects and the measurements at the LHC can be used to constrain the higher twist contributions.

\section{Conclusions}
\label{concl}

A study of the forward Drell--Yan cross sections in $pp$ collisions was performed in the $k_T$ factorization framework in the color dipole realizations. We assumed the LHC energy $\sqrt{S} = 14$~TeV and the forward kinematics --- the Feynman~$x$ of the DY pair, $x_F =0.05$. There are two important ingredients which influence the Drell--Yan process beyond the collinear approximation: the higher twist contributions and the transverse momenta of partons. At lower values of the boson invariant mass $M^2$ and at a very small $x$, both the effects are significant and of similar magnitude and may compete with each other. The disentanglement of those distinct contributions requires a careful analysis of the Drell--Yan structure functions over a broader range of kinematical parameters. The higher twist corrections are strongly suppressed at large process scales given by the Drell--Yan pair mass $M$ and/or the transverse momentum $p_T$. Thus in that kinematical region of $M \gg 10$~GeV and/or $p_T > 10$~GeV the higher twist effects may be safely neglected and the parton $k_T$ effects may be isolated. The results may be then used to provide necessary input to fit the parton transverse momentum distributions. A good and encouraging example is provided by a recent ATLAS measurement of the Lam--Tung relation breaking at the $Z^0$ boson peak \cite{ATLASZ0} which may serve as an excellent test for the gluon transverse momentum distributions \cite{MoSadSteLT}. With good understanding of the parton $k_T$ achieved, the higher twist effects may be probed in detail. Hence the observation of the higher twist effects should be possible at lower values of $M^2$, desirably at $M^2 < 100$~GeV$^2$, and very small $x < 10^{-5}$ of one of the partons, and at a low $p_T$. For optimal quality of the higher twist determination, the measurements of all the Drell--Yan structure functions should be performed, in a wide range of kinematical parameters. The $W_{TT}$ structure function and the Lam--Tung observable $A_{LT} = A_0 - A_2$ exhibit particularly high sensitivity to effects beyond the leading twist collinear approximation.

\section*{Acknowledgements}
Support of the Polish National Science Centre grants no.\ DEC-2014/13/B/ST2/02486 is gratefully acknowledged.
TS acknowledges support in the form of a scholarship of Marian Smoluchowski Research Consortium Matter Energy Future from KNOW funding.

\appendix

\section*{Appendix}
\section{Impact factors}
The Mellin transforms of leptonic impact factors were calculated in \cite{MoSadSte}. The results for the $q_T$-dependent impact factors and for the $q_T$-integrated impact factors are the following:
\subsection{The $q_T$-dependent impact factors}
\label{impact_facts_unint}
\begin{eqnarray}
\hat{\Phi}_{L} (q_T,s,z)&=& \frac{2}{z^2} \left\{ \frac{2 \Gamma^2(s+1) }{1+q_T^2/\eta^2_z} \ {}_2 F_1 \left (s+1,s+1,1,-\frac{q_T^2}{\eta^2_z} \right)
\right. \nonumber \\
& & - \left.
\Gamma(s+1) \Gamma(s+2) \ {}_2 F_1 \left (s+1,s+2,1,-\frac{q_T^2}{\eta^2_z} \right) \right\} ,
\label{sigmaLnonInt}\\
%\end{eqnarray}
%\begin{eqnarray}
\hat{\Phi}_{T} (q_T,s,z) &=&
\frac{1+(1-z)^2}{2z^2(1-z)}\Bigg\{
\frac{2 q_T^2/\eta^2_z}{1+q_T^2/\eta^2_z} \Gamma(s+1) \Gamma(s+2) \ {}_2 F_1 \left(s+1,s+2,2,-\frac{q_T^2}{\eta^2_z} \right) \Bigg. \nonumber \\
& & - \Gamma(s+1)^2
\left[
{}_2 F_1 \left(s+1,s+1,1,-\frac{q_T^2}{\eta^2_z} \right) \nonumber
\right.
\\
& & \Bigg. \left.
-(s+1)\ {}_2 F_1 \left (s+1,s+2,1,-\frac{q_T^2}{\eta^2_z} \right) \right] \Bigg\},
\label{sigmaTnonInt} \\
%\end{eqnarray}
%\begin{eqnarray}
\hat{\Phi}_{TT} (q_T,s,z) &=& \frac{1}{2 z^2}\left\{ \frac{2\pi }{\Gamma(1-s)\sin\pi s \ q_T^2/\eta^2_z} \left(1+\frac{q_T^2}{\eta^2_z} \right)^{-s-3} \Gamma(s+2)
\nonumber \right.\\
& & \left. \left[ \left(1+\frac{q_T^2}{\eta^2_z} \right)\left(1+\frac{q_T^2}{\eta^2_z}(s+2) \right) \ {}_2 F_1 \left (-s+1,s+1,1,\frac{q_T^2}{q_T^2+\eta^2_z} \right) \right. \right. \nonumber \\
& & \left. - \left(1+2\frac{q_T^2}{\eta^2_z}(s+1) \right) {}_2 F_1 \left (-s+1,s+2,1,\frac{q_T^2}{q_T^2+\eta^2_z} \right) \right]\nonumber \\
& & - \left. \frac{4q_T^2/\eta^2_z}{1+q_T^2/\eta^2_z} \Gamma(s+1) \Gamma(s+2) \ {}_2 F_1 \left (s+1,s+2,2,-\frac{q_T^2}{\eta^2_z} \right) \right\} ,
\label{sigmaTTnonInt} \nonumber\\
\end{eqnarray}
\begin{eqnarray}
\hat{\Phi}_{LT} (q_T,s,z)&=&\frac{2-z}{z^2 \sqrt{1-z}} \left\{ \pi \frac{q_T/\eta_z}{(1+q_T^2/\eta^2_z)^{s+2}} \ \frac{\Gamma(s+2)}{\Gamma(-s-1)\sin\pi s} \ {}_2 F_1 \left (-s,s+2,2,\frac{q_T^2}{q_T^2+\eta^2_z} \right) \right.
\nonumber \\
& & - \left. \frac{q_T/\eta_z}{1+q_T^2/\eta^2_z} \Gamma^2(s+1)\ \left[ \ {}_2 F_1 \left (s+1,s+1,1,-\frac{q_T^2}{\eta^2_z} \right) \right. \right. \nonumber \\
& & + \left. \left. (s+1) \ {}_2 F_1 \left (s+1,s+2,2,-\frac{q_T^2}{\eta^2_z} \right) \right] \right\},
\label{sigmaLTnonInt}
\end{eqnarray}
where $\eta^2_z =M^2(1-z)$.

\subsection{The $q_T$-integrated impact factors}
\label{impact_facts_int}
\begin{eqnarray}
H_L &=& \frac{\sqrt{\pi}\ \Gamma^3(s+1)}{\Gamma\left(s+\frac{3}{2}\right)}, \nonumber\\
H_T &=& \frac{\sqrt{\pi}\ \Gamma(s) \Gamma(s+1) \Gamma(s+2)}{4\Gamma\left(s+\frac{3}{2}\right)}, \nonumber\\
H_{TT} &=& \frac{\Gamma(s) \Gamma(s+1) \left[ 4^s\ \Gamma\left(s+\frac{3}{2}\right) -\sqrt{\pi}\ \Gamma(s+2) \right]}{2\Gamma\left(s+\frac{3}{2}\right)}, \nonumber\\
H_{LT} &=& -4^{s-1} \ \Gamma^2\left(s+\frac{1}{2} \right)+ \chi_1(s)+\chi_2(s),
\end{eqnarray}
where:
\begin{eqnarray}
\chi_1(s)&=&\int_0^\infty \frac{dt}{(1+t^2)^{1/2}} \int_0^\infty d\rho \ \rho^{2s+1} \sin(\rho t) K_0(\rho),%=\int_0^\infty d\rho \ \rho^{2s+1} \sin(\rho t) K_0(\rho)A(\rho)
\nonumber \\
\chi_2(s)&=&-\int_0^\infty \frac{dt}{(1+t^2)^{3/2}} \int_0^\infty d\rho \ \rho^{2s} \sin(\rho t) K_1(\rho).%=\int_0^\infty d\rho \ \rho^{2s+1} \sin(\rho t) K_1(\rho)A'(\rho)
\label{kappaDef}
\end{eqnarray}
\section{Coefficients of twist expansion}
\label{coeff_a_ex}
Here we present $a_{m}^{(2n)i}$ that were obtained by the series expansion of (\ref{hn2}). Coefficients for $m \geq 1$ are not presented as they are lengthy.
\begin{equation} \label{coefs}
\begin{array}{l}
a_{0}^{(2)T} =-\frac{M^4 (2-(2-z) z) \left(M^4 (1-z)^2+q_T^4\right)}{2 \left(q_T^2+M^2 (1-z)\right){}^4},\\
a_{0}^{(2)L} = -\frac{4 M^6 q_T^2 (1-z)^2 }{\left(q_T^2+M^2 (1-z)\right){}^4},\\
a_{0}^{(2)LT} =\frac{M^5 q_T (1-z) (2-z) \left(q_T^2-M^2 (1-z)\right)}{\left(q_T^2+M^2 (1-z)\right){}^4},\\
a_{0}^{(2)TT} = -\frac{2 M^6 q_T^2 (1-z)^2 }{\left(q_T^2+M^2 (1-z)\right){}^4}, \\
a_{0}^{(4)T} = -e^{-2 Y\bar{ \alpha} _s}\frac{M^6 z^2 (2-(2-z) z) \left(q_T^2-2 M^2 (1-z)\right) \left( M^2 (1-z)-(2+\sqrt{3}) q_T^2\right) \left( M^2 (1-z)- (2-\sqrt{3}) q_T^2 \right)}{\left(q_T^2+M^2 (1-z)\right){}^6},\\
a_{0}^{(4)L} = -e^{-2 Y\bar{ \alpha} _s}\frac{4 M^8 z^2 (1-z)^2 \left( M^2 (1-z)-(5+3\sqrt{2}) q_T^2 \right) \left( M^2 (1-z)- (5-3\sqrt{2}) q_T^2 \right)}{\left(q_T^2+M^2(1-z)\right){}^6},\\
a_{0}^{(4)LT} = e^{-2 Y\bar{ \alpha} _s}\frac{2 M^7 q_T z^2 (1-z) (2-z) \left(5 M^2 (1-z)- q_T^2 \right) \left( M^2 (1-z)- 2 q_T^2\right)}{ \left(q_T^2+M^2 (1-z)\right){}^6},\\
a_{0}^{(4)TT} =-e^{-2 Y\bar{ \alpha} _s}\frac{12 M^8 q_T^2 z^2 (1-z)^2 \left(q_T^2-2 M^2 (1-z)\right)}{\left(q_T^2+M^2 (1-z)\right){}^6}. \\
\end{array}
\end{equation}
Notice that the twist~4 terms are suppressed by the $e^{-2 Y \bar{\alpha}_s}$ factor. In fact, it turns out that contribution from the higher twist terms in the BFKL model is negligible as is seen e.g.\ 
from Fig.\ \ref{coefficientsW_ratios}.

\section{Twist expansion for $\tilde{W}_{T}$ in BFKL model}
\label{twist_T}

The twist expansion of the structure function $\tilde{W}_{T}$ in the BFKL model requires a careful treatment of the $z$ integration at the singular $z\to 1$ limit. It is performed following the procedure proposed in Ref.\ \cite{GBLS}. Hence we rewrite (\ref{Wi2n}) for $W^{(2)}_{T}$ in the following form:
\begin{eqnarray}
\label{WT2}
\tilde{W}_{T} ^{(2)} &=& -\sigma_{0}' e^{-\tilde{t}} \int_{x_F}^1 dz\int_0^{2\pi} d \theta \ \left[1+(1-z)^2 \right] \wp(x_F/z) \ \mathrm{exp} \left( - \epsilon e^{i \theta} \ \mathrm{ln}z^2 \right) \nonumber \\
& &\times \ \ \tilde{h}_T^{(2)} \left( \frac{1}{1-z} \right)^{1-\epsilon \, \mathrm{exp} \, i \theta}
\mathrm{exp} \left( \epsilon e^{i \theta} \ \tilde{t} + \bar{\alpha}_s Y \frac{1}{ \epsilon e^{i \theta} } \right),
\end{eqnarray}
and then the first line of this expression as:
\begin{eqnarray}
\label{expansion_in_z=1}
\left[1+(1-z)^2 \right] && \wp(x_F/z) \ \mathrm{exp} \left( - \epsilon e^{i \theta} \ \mathrm{ln}z^2 \right) \nonumber \\
&&=\wp(x_F) +\left\{ \left[1+(1-z)^2 \right] \wp(x_F/z) \ \mathrm{exp} \left( - \epsilon e^{i \theta} \ \mathrm{ln}z^2 \right) - \wp(x_F) \right\}.
\end{eqnarray}
After the integration of the first term, $\wp(x_F)$, one has:
\begin{eqnarray}
\label{WT2'}
\tilde{W}_{T} ^{(2)'} &=& -\sigma_{0}' \left( \frac{\bar Q_0^2}{4M^2} \right) \wp(x_F) \sum_{m=0}^{\infty} \tilde{a}_m^{(2)T}(x_F)
\left( \frac{\bar{\alpha}_s Y}{ \textrm{ln} ( 4M^2/\bar Q_0^2 ) } \right)^{\frac{m-1}{2}} \nonumber \\
& &\times \ I_{|m-1|} \left( 2\sqrt{\bar{\alpha}_s Y \textrm{ln} \frac{4M^2}{\bar Q_0^2} }\, \right).
\end{eqnarray}

The term $\left\{ \ldots \right\}$ from (\ref{expansion_in_z=1}) is proportional to $(1-z)$ since l.h.s. of (\ref{expansion_in_z=1}) is analytical around $z=1$. So after substituting this term into (\ref{WT2}) we get a convergent integral over $z$:
\begin{eqnarray}
\label{WT2''}
\tilde{W}_{T} ^{(2)''} &=& -\sigma_{0}' \left( \frac{\bar Q_0^2}{4M^2} \right) \sum_{m=0}^{\infty} \int_{x_F}^1 \frac{dz}{1-z}\ \tilde{a}_m^{(2)T}(z) \nonumber \\
&\times &\ \left\{ \left( 1+(1-z)^2 \right) \wp(x_F/z) \left( \bar{\alpha}_s Y/ \textrm{ln} \frac{4M^2}{z^2 \bar Q_0^2} \right)^{\frac{m}{2}} I_{|m|} \left( 2\sqrt{\bar{\alpha}_s Y \ \textrm{ln} \frac{4M^2}{z^2 \bar Q_0^2} }\, \right) \right. \nonumber \\
& & -\ \left. \wp(x_F) \left( \bar{\alpha}_s Y/ \textrm{ln} \frac{4M^2}{\bar Q_0^2} \right)^{\frac{m}{2}} I_{|m|} \left( 2\sqrt{\bar{\alpha}_s Y \ \textrm{ln} \frac{4M^2}{\bar Q_0^2} }\, \right) \right\}.
\end{eqnarray}
Twist 2 of $\tilde{W}_T$ is a sum of $\tilde{W}_{T} ^{(2)'}$ (\ref{WT2'}) and $\tilde{W}_{T} ^{(2)''}$ (\ref{WT2''}).

Finally, let us give first two $\tilde{a}_m^{(2)T}$ coefficients:
\begin{eqnarray}
\tilde{a}_0^{(2)T}(z) &=& -\frac{2}{3}, \ \ \ \tilde{a}_1^{(2)T}(z) = \frac{1}{3} \left[ 3 -4 \gamma_E -2 \ \textrm{ln} (1-z) - 2 \psi(5/2) \right].
\end{eqnarray}

Twist 4 for $\tilde{W}_L$, $\tilde{W}_{TT}$ and $\tilde{W}_{LT}$ could be obtained similarly as $\tilde{W}_{T} ^{(2)}$. Twist 4 for $\tilde{W}_{T}$ is more complicated because $\int^1 dz \ 1/(1-z)^2$ divergence occurs, see \cite{GBLS} for the details of computation in GBW model.

\end{document}